\documentclass[11pt,table]{article}

\usepackage[preprint]{acl}

\usepackage{times}
\usepackage{latexsym}

\usepackage[T1]{fontenc}

\usepackage[utf8]{inputenc}

\usepackage{microtype}
\usepackage{enumitem}
\usepackage{inconsolata}

\usepackage{graphicx}
\usepackage[table]{xcolor}

\usepackage[most]{tcolorbox}

\usepackage{booktabs} 
\usepackage{multirow}
\usepackage{url}
\usepackage{relsize}
\usepackage{amssymb}    
\usepackage{amsmath}    
\usepackage{amsthm}
\usepackage{xcolor}

\usepackage{cuted}
\usepackage[most]{tcolorbox}
\definecolor{boxblue}{RGB}{45, 75, 115}    
\definecolor{bgblue}{RGB}{252, 253, 255}   

\newtcblisting[auto counter, number within=section]{promptbox}[2][]{
    enhanced,
    breakable,
    listing only,
    colback=bgblue,
    colframe=boxblue,
    boxrule=0.5pt,
    fonttitle=\small\bfseries\sffamily,
    coltitle=white,
    colbacktitle=boxblue,
    attach boxed title to top left={yshift=-1mm, xshift=2mm},
    boxed title style={sharp corners, boxrule=0pt, left=2pt, right=3pt, top=1pt, bottom=1pt},
    title={Prompt \thetcbcounter: #2},
    label={#1}, 
    listing options={
        basicstyle=\small\ttfamily,
        breaklines=true,        
        breakindent=0pt,       
        breakatwhitespace=true,
        keepspaces=true,
        columns=flexible,       
    },
    left=6pt, right=6pt, top=6pt, bottom=6pt,
}

\usepackage{colortbl}
\newcommand{\fboxtag}[1]{%
  \begingroup
  \setlength{\fboxsep}{1.5pt}
  \fbox{\textbf{\scriptsize\rmfamily #1}}
  \endgroup
}

\newenvironment{CaseStudyTable}[2]{%
  \begin{table*}[h]
  \centering
  \caption{#1}\label{#2}
  \fontsize{9pt}{10pt}\rmfamily\selectfont     
  \begin{tabular}{p{0.98\linewidth}}
  \midrule
}{%
  \midrule
  \end{tabular}
  \end{table*}
}

\usepackage{CJKutf8}  

\usepackage{algorithm}
\usepackage{algpseudocode}
\usepackage{amsmath}      
\usepackage{amssymb}      
\usepackage[capitalize]{cleveref}

\title{GroupRank: A Groupwise Paradigm for Effective and Efficient Passage Reranking with LLMs}

\author{
 \textbf{Meixiu Long\textsuperscript{1,2}\thanks{Equal contribution.}\thanks{Work done during the internship at Ant Group.}},
 \textbf{Duolin Sun\textsuperscript{2}}\footnotemark[1],
 \textbf{Dan Yang\textsuperscript{2}\thanks{Corresponding authors.}},
  \textbf{Yihan Jiao\textsuperscript{2}},
 \textbf{Lei Liu\textsuperscript{2}},
 \textbf{Jiahai Wang\textsuperscript{1}\footnotemark[3]},
 \textbf{BinBin Hu\textsuperscript{2}},
 \textbf{Yue Shen\textsuperscript{2}},
\\
 \textbf{Jie Feng\textsuperscript{2}},
  \textbf{Zhehao Tan\textsuperscript{2}},
    \textbf{Junjie Wang\textsuperscript{2}},
  \textbf{Lianzhen Zhong\textsuperscript{2}},
   \textbf{Jian Wang\textsuperscript{2}},
 \textbf{Peng Wei\textsuperscript{2}},
 \textbf{Jinjie Gu\textsuperscript{2}}
\\
 \textsuperscript{1}School of Computer Science and Engineering, Sun Yat-sen University,
 \textsuperscript{2}Ant Group
\\
 \small{
   \textbf{Correspondence:} {longmx7@mail2.sysu.edu.cn,\{sunduolin.sdl, luoyin.yd\}@antgroup.com},wangjiah@mail.sysu.edu.cn}
}

\begin{document}
\maketitle
\begin{abstract}
Large Language Models (LLMs) have emerged as powerful tools for passage reranking in information retrieval, leveraging their superior reasoning capabilities to address the limitations of conventional models on complex queries. However, current LLM-based reranking paradigms are fundamentally constrained by an efficiency-accuracy trade-off: (1) pointwise methods are efficient but ignore inter-document comparison, yielding suboptimal accuracy; (2) listwise methods capture global context but suffer from context-window constraints and prohibitive inference latency. To address these issues, we propose \textbf{GroupRank}, a novel paradigm that balances flexibility and context awareness. 
To unlock the full potential of groupwise reranking, we propose an answer-free data synthesis pipeline that fuses local pointwise signals with global listwise rankings. 
These samples facilitate supervised fine-tuning and reinforcement learning, with the latter guided by a specialized group-ranking reward comprising ranking-utility and group-alignment. These complementary components synergistically optimize document ordering and score calibration to reflect intrinsic query-document relevance.
Experimental results show GroupRank achieves a state-of-the-art 65.2 NDCG@10 on BRIGHT and surpasses baselines by 2.1 points on R2MED, while delivering a 6.4$\times$ inference speedup. The code is available at \url{https://github.com/AQ-MedAI/Diver/tree/main/Reranker/GroupRank}.

\end{abstract}

\begin{figure}[htbp]
    \centering
    \includegraphics[width=1\linewidth]{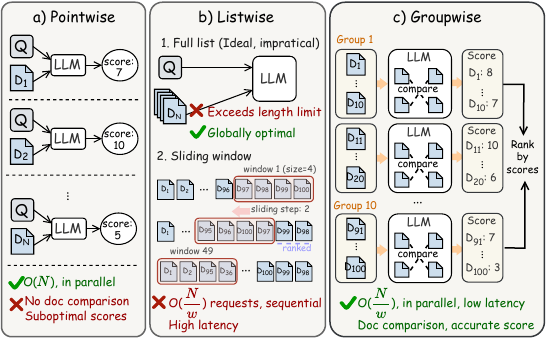}
    \caption{Comparison of LLM reranking paradigms. Unlike suboptimal pointwise and inefficient listwise methods, our groupwise approach performs parallel cross-document comparisons within groups, achieving high accuracy with superior efficiency in practice.} 
    \label{fig:Reranker-intro}
\end{figure}

\section{Introduction}
Passage reranking is a crucial component in Information Retrieval (IR), reordering relevant passages for downstream tasks such as open-domain question answering~\cite{rag3,metrag_2024} and web search~\citep{25icde_web_search_intro}. Existing rankers excel in lexical-matching scenarios~\citep{16nips_msmarco}, while they struggle with complex queries requiring deep reasoning, as exemplified by benchmarks like BRIGHT~\citep{25iclr_bright}. For instance, a programmer seeking documentation for a specific bug must understand the algorithmic logic involved rather than merely matching variable names. To capture such implicit relevance beyond shallow semantics, Large Language Models (LLMs)~\citep{Qwen3_report,DeepSeek-R1} are introduced, empowering reranking models with their superior reasoning capabilities.

Two widely adopted LLM-based reranking paradigms are the pointwise and listwise approaches, as illustrated in \cref{fig:Reranker-intro}. Pointwise methods predict a relevance score for each query-document pair independently. However, the assigned scores are often inaccurate due to the absence of cross-document comparison, leading to suboptimal ranking. By contrast, listwise methods input multiple documents simultaneously to capture global relevance. While promising, feeding a large candidate pool (\textit{e.g.}, top-100 passages) typically exceeds the context window constraints of LLMs. Sliding window strategies mitigate this by processing the list in sequential batches, but the iterative mechanism incurs prohibitive latency and computational overhead, rendering such methods impractical for real-world deployment. This raises a critical question: \textit{Can we achieve the high effectiveness of listwise reranking while maintaining the efficiency of pointwise methods?}

To achieve this, we propose a novel \underline{group}wise \underline{rank}ing (\textbf{GroupRank}) method illustrated in \cref{fig:Reranker-intro}. It partitions the candidate documents into small, independent groups (\textit{e.g.}, 10 per group) for joint scoring. This design enables comparison across documents, capturing relative relevance signals missed by isolated pointwise scoring. Furthermore, decomposing the list naturally avoids context window limits. Unlike sequential listwise models, the independent groups enable fully parallel processing, followed by score aggregation to obtain the final ranking. Consequently, it substantially reduces LLM inference calls and achieves lower latency than even pointwise methods.

To fully unlock the potential of GroupRank, we develop a training framework that includes an \textbf{answer-free synthesis pipeline} and a specialized reward for groupwise reranking. Specifically, we first present a synthesis pipeline to alleviate data scarcity without relying on ground-truth answers. By integrating local pointwise signals with global listwise rankings, we generate 14k high-quality samples that capture both individual and relative relevance. 
Training involves Supervised Fine-Tuning~(SFT) for format alignment, followed by Reinforcement Learning~(RL) to stimulate intra-group comparison for more precise scoring. The RL stage is guided by a specialized \textbf{Group-Ranking Reward}, comprising \textbf{ranking-utility} for positional precision and \textbf{group-alignment} for distribution regularization. These complementary views synergistically guide the model to achieve superior reranking effectiveness while producing calibrated scores that reflect the intrinsic query-document relevance.

Experimental results show that GroupRank achieves state-of-the-art (SOTA) performance, reaching 65.2 and 52.3 NDCG@10 on BRIGHT and R2MED, respectively, while also excelling on semantic datasets. Our main contributions are:
\begin{itemize}[leftmargin=*]
    \item We propose GroupRank, a groupwise reranking paradigm that balances listwise effectiveness with pointwise efficiency, utilizing parallel cross-document comparisons to yield superior ranking with low latency.
    \item An answer-free pipeline is proposed to synthesize high-quality data for two-stage training. For the RL stage, a \textbf{Group-Ranking reward} designed for the groupwise paradigm optimizes both ranking positions and relevance score magnitudes.
    \item GroupRank outperforms leading baselines by up to 2.1 NDCG@10 on reasoning-intensive tasks and 3.0 on semantic benchmarks, while delivering 2.4-6.4$\times$ faster inference.
\end{itemize}

\section{Related Work}
\paragraph{LLM-based Ranking}
LLMs have revolutionized reranking by capturing implicit relevance beyond surface-level semantics. Existing methods are generally categorized into \textit{pointwise}, \textit{pairwise}, and \textit{listwise} (setwise) paradigms. Pointwise rerankers~\citep{ERank,retro_ranker,Rank1} evaluate query-document pairs independently, offering $O(N)$ efficiency suitable for large-scale systems. However, they often yield suboptimal results due to the lack of cross-document comparison. Pairwise approaches~\citep{pairwise_ranking} address this by comparing document pairs to improve accuracy but incur prohibitive quadratic computational costs. Listwise methods~\citep{ReasonRank,RankZephyr,Rank-R1,listwise_FIRST} process candidate lists jointly to capture global relevance patterns, achieving superior performance. Nevertheless, the context window constraints of LLMs necessitate sequential sliding window strategies~\citep{sliding_window}, resulting in significant latency that hinders practical deployment.
To address these limitations, this paper proposes a groupwise framework to balance the efficiency of pointwise models with the accuracy of listwise paradigms.

\begin{figure*}[h]
    \centering
    \includegraphics[width=0.9\textwidth]{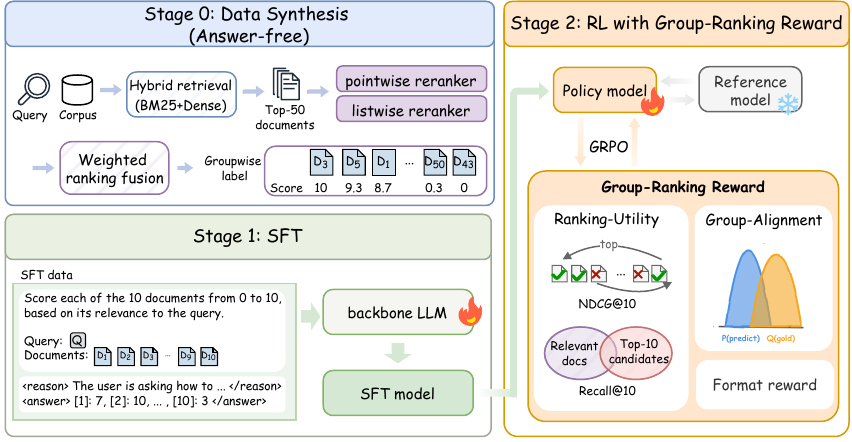}
    \caption{\textbf{Overview of GroupRank.} The pipeline starts by synthesizing training samples without ground-truth answers, followed by supervised fine-tuning to align groupwise formats and stabilize training. Finally, reinforcement learning with Group-Ranking reward is employed to optimize ranking accuracy and calibrate score distributions. }
    \label{fig:overview_grouprank}
\end{figure*}

\paragraph{Training Strategies for Ranking}
Early research utilized zero-shot prompting~\citep{JudgeRank}, but high costs and limited adaptability shifted the focus toward specialized training. Recent efforts~\citep{RankZephyr,RankT5,Rank1} utilize distillation and SFT to handle complex reasoning tasks, such as those in the BRIGHT benchmark. For instance, Rank-K~\citep{Rank-K} distills reasoning chains from Large Reasoning Models like DeepSeek-R1~\cite{DeepSeek-R1} to enhance listwise capabilities. Subsequent works~\citep{Rank-R1,ERank,ReasonRank} have incorporated RL for further improvement. ReasonRank~\citep{ReasonRank} addresses data scarcity through automated synthesis, but requires ground-truth answers to guide the generation process. Similarly, ERANK~\cite{ERank} trains a pointwise reranker using listwise-derived RL signals, enhancing relevance discrimination through fine-grained scoring.

\section{Methodology}
In this section, we first introduce how to generate high-quality training samples without external labels. Then, SFT followed by a modified GRPO algorithm is to establish foundational groupwise ranking capabilities. Overall framework is illustrated in~\cref{fig:overview_grouprank}.

\subsection{Definition of Groupwise Ranking}
Given a query $q$ and a set of $N$ candidate documents $\mathcal{D} = \{d_1, d_2, \dots, d_N\}$, the groupwise paradigm partitions $\mathcal{D}$ into $M = \lceil \frac{N}{c} \rceil$ disjoint groups $\{G_1, G_2, \dots, G_M\}$, where each group contains $c$ documents. For each group $G_i$, a scoring function $f_\theta$ simultaneously evaluates the relevance scores for all documents within $G_i$:
\begin{equation}
\resizebox{0.85\linewidth}{!}{
$\mathbf{s}_i = f_\theta(q, G_i) = [s_{i,1}, s_{i,2}, \dots, s_{i,c}] \in \mathbb{R}^c,$
}
\label{eq:groupwise_paradigm}
\end{equation}
where $s_{i,j}$ represents the relevance score of the $j$-th document in $G_i$. All scores are aggregated into a final set $\mathcal{S}_q = \bigcup_{i=1}^M \mathbf{s}_i$, and the documents in $\mathcal{D}$ are sorted in descending order of these scores to obtain the ranked list. Moreover, this paradigm supports test-time scaling (TTS)~\citep{retro_ranker} by evaluating a document within multiple groups. Averaging these scores integrates diverse comparative perspectives, resulting in more precise final rankings, as demonstrated in \cref{sec:exp-further_enhancement}.

\subsection{Data Synthesis Pipeline}
To address the scarcity of high-quality labels in reasoning-intensive scenarios, we propose an answer-free data synthesis pipeline. Unlike existing methods that require query-specific gold answers for label generation~\citep{ReasonRank} or simply distill ranking results generated by LLMs~\cite{ERank,retro_ranker}, our approach produces training samples without golden answers. Instead, we integrate local and global relevance signals from pointwise and listwise models to generate high-quality pseudo-labels.

\paragraph{Query and Candidate Construction}
The effectiveness of reranking models depends heavily on the diversity of training data and the quality of negative samples. If negative samples are overly simplistic, the model often fails to develop the fine-grained discriminative capabilities required for complex reasoning tasks. To ensure generalization across various scenarios, we curate a multi-domain query corpus containing approximately 13,600 samples. This corpus includes 13,000 queries for general logical reasoning~\citep{ReasonRank}, 1,000 queries from the in-house medical data, and 700 programming queries obtained from the xCodeEval dataset~\citep{xCodeEval}.

For each query $q$, we employ a hybrid retrieval strategy to collect candidate documents. This approach leverages the complementary strengths of sparse and dense retrieval. BM25~\citep{BM25_Pyserini} excels at keyword matching, while dense retrieval models~\citep{DIVER} are effective at identifying semantic relevance. We integrate these two signals using a weighted sum of normalized scores:
$ S_{\text{hybrid}}(q, d) = \alpha \cdot \text{norm}(S_{\text{BM25}}(q, d)) + (1 - \alpha) \cdot \text{norm}(S_{\text{dense}}(q, d))$,
where $S_{\text{BM25}}(q, d)$ and $S_{\text{dense}}(q, d)$ denote the raw scores from the respective models. The $\text{norm}(\cdot)$ function maps raw scores into a unified range between 0 and 1, and $\alpha=0.5$.

To construct a challenging candidate pool, we select the top 10 documents as potential positives and randomly sample an additional 40 documents from the ranking range of 11 to 200 as hard negatives. This strategy ensures the pool contains both highly relevant documents and misleading samples that are difficult to distinguish. This process yields a candidate pool of 50 documents per query. Such an extensive pool provides the necessary flexibility to construct training sub-lists of various lengths and difficulty levels, which is essential for the subsequent multi-scale training. 

\paragraph{Ranking Pseudo-Labeling}
In the absence of ground-truth answers, relying on a single teacher model often introduces inherent biases. 
Pointwise scoring lacks cross-document comparison, while listwise ranking fails to quantify the specific magnitude of relevance. To generate high-quality ranking supervision, we present a dual-teacher framework that synergizes local pointwise and global listwise perspectives. For the pointwise teacher, we employ Qwen3-235B-instruct-2507~\citep{Qwen3_report}, a capable open-source model that provides reliable relevance scores on a per-document basis. For the listwise teacher, Gemini-2.5-Pro~\citep{Gemini-2.5} is utilized for its ability to process ultra-long contexts up to 1M tokens, enabling it to jointly evaluate and rank the entire candidate set of 50 documents.

The pointwise teacher assigns a score $S_p \in [0, 10]$ to each query-document pair, yielding a ranked list $L_p$ after sorting. The listwise teacher directly produces a global ranking $L_l$ of all 50 documents.
To integrate these complementary signals, we map both rankings into a continuous score space using a negative logarithmic transformation. This transformation effectively converts ordinal positions into relevance scores while amplifying the discrimination between top-ranked candidates. Following the fusion strategy presented in \citep{DIVER}, we calculate the final supervision score $S_{\text{label}}$ by aggregating the transformed ranks:
\begin{equation}
\resizebox{0.85\linewidth}{!}{
  $S_{\text{label}}(d) = -\beta \log(L_p(d)) - (1-\beta) \log(L_l(d)),$
}
\label{eq:fusion_score}
\end{equation}
where $L_p(d)$ and $L_l(d)$ denote the rank positions of document $d$ in the pointwise and listwise results, respectively. Hyperparameter $\beta$ balancing the two signals is set to $0.5$. The final high-quality ranking $L_{\text{label}}$ is generated by sorting documents in descending order of $S_{\text{label}}$. 
This strategy effectively integrates pointwise relevance (the local view) with listwise relationship (the global view), as empirically validated in \cref{sec:app-Dual-Teacher-Fusion}.

To optimize generation efficiency, we perform dual-teacher ranking only once for the 50-document pool and then apply a \textit{multi-scale sampling mechanism}. This approach generates 14k training samples with group sizes ranging from 5 to 20, ensuring model robustness across various candidate scales while significantly reducing API computational overhead.

\subsection{Two-stage Training for GroupRank}

\paragraph{Stage-1: Cold Start SFT} 
This stage aligns base models~(\textit{e.g.}, Qwen2.5-7B) with the groupwise format in Eq.(1). SFT instills the capability to generate reasoning chains within \texttt{<reason>} tags and JSON-formatted scores within \texttt{<answer>} tags. We optimize the parameters $\theta$ using the standard objective on a synthetic dataset $\mathcal{T}$:
\begin{equation}
\mathcal{L}_{\text{SFT}}(\theta) = -\sum_{(x,y) \in \mathcal{T}} \sum_{t=1}^{|y|} \log P_{\theta}(y_t \mid x, y_{<t}),
\end{equation}
where $x$ and $y$ represent the input prompt and target sequence, respectively. This procedure establishes foundational groupwise reranking proficiency. Further details are provided in \cref{sec:app-exp-setting}.

\paragraph{Stage-2: RL with Group-Ranking Reward} 
Reinforcement learning facilitates reasoning chain exploration, which is essential for GroupRank to discern subtle ranking differences through intra-group cross-document comparisons. To achieve this, we design a customized Group-Ranking Reward specifically for the groupwise paradigm and employ the Group Relative Policy Optimization~(GRPO) algorithm for efficient optimization.

\paragraph{Group-Ranking Reward} 
The reward signal serves as the optimization objective that directly guides the policy model. We design a comprehensive Group-Ranking Reward tailored for the groupwise ranking paradigm, incorporating ranking quality and score distribution.
First, the \textbf{Ranking-Utility Reward} optimizes the core ranking performance by combining NDCG@10 and Recall@10. NDCG@10 provides positional sensitivity by assigning higher weights to relevant documents at the top of the list, whereas Recall@10 measures relevance coverage to ensure that high-relevance items are successfully captured within the top results.
Second, the \textbf{Group-Alignment Reward} employs Jensen-Shannon (JS) Divergence to regularize the score distribution. This symmetric and stable metric aligns predicted scores with a synthetic gold standard to preserve magnitude information. It effectively prevents reward hacking, where the model outputs extreme probabilities to maximize utility metrics without reflecting actual relevance levels. It ensures that predicted scores remain calibrated of actual query-document relevance rather than merely justifying a relative order.

Beyond the ranking performance, we implement a Format Reward as a structural constraint to validate the presence of <reason> and <answer> tags as well as the adherence to the specified JSON format. To prioritize structural integrity during training, $R$ is formulated as:
\begin{equation}
\label{eq:reward_function}
\resizebox{0.85\linewidth}{!}{
$ R = 
\begin{cases}
R_{G}, & \text{Correct tags and JSON format,} \\
-0.1, & \text{Correct tags but invalid JSON,} \\
-0.5, & \text{Otherwise.} 
\end{cases} $
}
\end{equation}
The Group-Ranking Reward $R_{G}$ is defined as the weighted sum of utility and alignment components:
\begin{equation} 
\label{eq:total_reward}
\resizebox{0.85\linewidth}{!}{
$R_{G} = \text{NDCG@10} + \gamma \cdot \text{Recall@10} + \epsilon \cdot (1 - \text{JS}), $
}
\end{equation}
\begin{equation}\text{JS} = \frac{1}{2} D_{\text{KL}}(P \parallel M) + \frac{1}{2} D_{\text{KL}}(Q \parallel M),
\end{equation}
where $P$ and $Q$ are the predicted and gold score distributions, and $\gamma, \epsilon$ are scaling parameters. $M = \frac{1}{2}(P + Q)$ represents the average distribution and $D_{\text{KL}}$ denotes the Kullback-Leibler divergence ~\citep{Kullback-Leibler-divergence}. The term $(1 - \text{JS})$ transforms the divergence into a positive similarity measure within $[0, 1]$ to encourage distribution alignment. This multi-dimensional objective ensures superior ranking precision while maintaining calibrated score magnitudes.

We optimize the policy network $\pi_{\theta}$ using GRPO. Both $\pi_{\theta}$ and the reference model $\pi_{\text{ref}}$ are initialized from the SFT-tuned reranker. For each input $x$, we sample $G$ outputs $\{y_1, \dots, y_G\}$ from the old policy $\pi_{\theta_{\text{old}}}$. Each sequence $y_i$ receives a reward $R_i$ based on the defined reward, which is then normalized within the group to compute the relative advantage $\hat{A}_{i,t}$. The objective maximizes a clipped surrogate loss with a KL penalty:
\begin{equation}
\resizebox{0.9\linewidth}{!}{ $
\begin{aligned}
\mathcal{J}_{\text{GRPO}}&(\theta) = \frac{1}{G} \sum_{i=1}^{G} \frac{1}{|y_i|} \sum_{t=1}^{|y_i|} \min \left(\rho_{i,t}(\theta) \hat{A}_{i,t}, \right. \\
&\quad \left. \text{clip} \left( \rho_{i,t}(\theta), 1 - \epsilon, 1 + \epsilon \right) \hat{A}_{i,t} \right) - \phi D_{\text{KL}}(\pi_{\theta} \| \pi_{\text{ref}}), \\
\rho_{i,t}(\theta) &= \frac{\pi_\theta(y_{i,t} \mid x, y_{i,<t})}{\pi_{\theta_{\text{old}}}(y_{i,t} \mid x, y_{i,<t})},
\end{aligned}
$}
\end{equation}
where $\rho_{i,t}(\theta)$ is the importance sampling ratio and $\phi$ is the KL penalty coefficient. $D_{\text{KL}}(\pi_{\theta} \| \pi_{\text{ref}})$ regularizes the policy towards the reference model. This iterative process enables GroupRank to refine its ranking precision through self-improvement while maintaining foundational model stability.

\section{Experiments}
\subsection{Experimental Setup}
\paragraph{Benchmarks} 
Experiments are conducted on two reasoning-intensive benchmarks, BRIGHT~\citep{25iclr_bright} and R2MED~\citep{r2med}, alongside the traditional semantic retrieval benchmark BEIR~\citep{beir} for comprehensive evaluation. Due to space constraints, benchmark and experimental details are deferred to~\cref{sec:app-experiments}.

\paragraph{Baselines} 
We compare GroupRank against representative baselines from two reranking paradigms. Pointwise rerankers: RankT5~\citep{RankT5}, ERank~\citep{ERank}. Listwise rerankers: FIRST~\citep{listwise_FIRST}, RankZephyr~\citep{RankZephyr}, Rank-R1~\cite{Rank-R1}, Rank-K~\cite{Rank-K}, and ReasonRank~\citep{ReasonRank}. We also evaluate Qwen2.5-32B-Instruct~\citep{qwen25}, Qwen3-32B~\citep{Qwen3_report}, and DeepSeek-V3.1\footnote{\url{https://huggingface.co/deepseek-ai/DeepSeek-V3.1}} in zero-shot groupwise settings using GroupRank’s prompt.

\begin{table*}[t]
    \centering
    \caption{NDCG@10 on the BRIGHT benchmark. Models rerank the top-100 passages retrieved by DIVER-Retriever-4B using GPT4-rewritten queries. Best and second-best results are \textbf{bolded} and \underline{underlined}.}
    \label{tab:result_bright}
    \footnotesize 
    \setlength{\tabcolsep}{4.2pt} 
    \resizebox{\textwidth}{!}{
    \begin{tabular}{lllcccccccccccc}
        \toprule
        \multirow{2}{*}{\textbf{Paradigm}} & \multirow{2}{*}{\textbf{Models}} & \multirow{2}{*}{\textbf{Avg.}} & \multicolumn{7}{c}{\textbf{StackExchange}} & \multicolumn{2}{c}{\textbf{Coding}} & \multicolumn{3}{c}{\textbf{Theorem-based}} \\
        \cmidrule(lr){4-10} \cmidrule(lr){11-12} \cmidrule(lr){13-15}
        & & & \textbf{Bio.} & \textbf{Earth.} & \textbf{Econ.} & \textbf{Psy.} & \textbf{Rob.} & \textbf{Stack.} & \textbf{Sus.} & \textbf{Leet.} & \textbf{Pony} & \textbf{AoPS} & \textbf{TheoQ.} & \textbf{TheoT.} \\
        \midrule
        Retriever & Diver-Retriever-4B & 32.9 & 52.5 & 53.6 & 33.8 & 45.2 & 28.4 & 30.4 & 35.0 & 13.0 & 14.7 & 9.8 & \textbf{42.5} & 36.3 \\ 
        \midrule
        \multirow{3}{*}{Pointwise} & RankT5-3B & 15.7 & 33.0 & 22.8 & 18.9 & 8.6 & 2.2 & 10.0 & 23.9 & 11.9 & 16.9 & 8.9 & 19.6 & 11.7 \\
        & ERank-4B & 31.8 & 48.2 & 53.0 & 31.0 & 44.5 & 22.4 & 24.8 & 33.7 & 9.7 & 18.2 & 11.1 & 38.2 & \underline{46.6} \\
        & ERank-32B & 36.2 & \underline{55.9} & \underline{55.1} & 35.2 & 44.6 & 34.1 & \underline{34.5} & 41.2 & 14.1 & \textbf{24.4} & \underline{12.2} & 36.7 & 45.9 \\
        \midrule
        \multirow{7}{*}{Listwise} 
        & FIRST-7B & 20.6 & 35.2 & 26.2 & 21.8 & 27.5 & 17.9 & 19.2 & 20.4 & 10.7 & 21.2 & 6.2 & 20.2 & 20.8 \\
        & RankZephyr-7B & 22.5 & 42.6 & 19.5 & 18.7 & 29.8 & 13.8 & 13.3 & 29.2 & 12.9 & 19.3 & 7.4 & 31.1 & 31.6 \\ 
        & Rank-R1-7B & 21.4 & 40.9 & 31.4 & 23.2 & 32.0 & 19.8 & 11.2 & 35.5 & 3.8 & 6.6 & 5.0 & 17.2 & 30.1 \\
        & Rank-R1-14B & 30.8 & 49.6 & 41.2 & 27.6 & 40.0 & 28.5 & 28.2 & 43.8 & 7.0 & 18.5 & 7.9 & 34.8 & 43.1 \\
        & Rank-K-32B & 33.3 & 51.1 & 42.3 & 33.0 & 44.5 & 33.1 & 28.4 & 41.6 & 12.8 & \underline{21.6} & 8.5 & 39.3 & 43.2 \\
        & ReasonRank-7B & 32.5 & 51.6 & 43.4 & 32.4 & 44.0 & 31.0 & 25.6 & 39.8 & \underline{15.4} & 20.1 & 7.0 & 38.9 & 40.7 \\
        & ReasonRank-32B & 35.6 & 53.9 & 47.6 & 36.3 & \textbf{52.6} & 36.5 & 34.2 & \underline{44.5} & 15.2 & 14.8 & 5.5 & 40.6 & 45.3 \\ 
        \midrule
        & Qwen2.5-32B-Instruct & 32.6 & 51.6 & 46.4 & 31.6 & 43.3 & 26.8 & 32.5 & 40.7 & 14.0 & 18.0 & 8.8 & 33.4 & 44.2 \\
        & Qwen3-32B (No-think) & 36.2 & \textbf{59.4} & 53.7 & \underline{38.7} & 49.6 & 35.6 & 33.4 & 43.7 & 13.6 & 13.8 & 9.5 & 38.9 & 44.4 \\
        & Qwen3-32B (Think) & 36.3 & 54.6 & 54.1 & 38.0 & 48.8 & 37.6 & 35.3 & 43.6 & 12.2 & 15.0 & 11.0 & 39.6 & 45.7 \\
        & DeepSeek-V3.1-671B & \underline{37.1} & 56.7 & 54.5 & 36.9 & 49.2 & \underline{38.9} & \underline{36.2} & \textbf{47.1} & 13.7 & 16.6 & 9.9 & 40.2 & 45.7 \\
        \rowcolor[gray]{0.94}
        & GroupRank-7B & 34.3 & 52.7 & 51.0 & 33.8 & 44.5 & 32.1 & 33.9 & 38.1 & \textbf{16.3} & 17.3 & 8.7 & \underline{40.7} & 42.4 \\
        \rowcolor[gray]{0.94}
        \multirow{-6}{*}{Groupwise} & GroupRank-32B & \textbf{38.0} & \underline{59.0} & \textbf{57.5} & \textbf{39.2} & \underline{50.0} & \textbf{39.1} & \textbf{39.0} & 42.7 & 14.3 & 14.9 & \textbf{12.6} & 39.0 & \textbf{48.8} \\
        \bottomrule
    \end{tabular}}
\end{table*}

\begin{table*}[t]
    \centering 
    \caption{NDCG@10 on the R2MED benchmark. Models rerank the top-100 passages retrieved with original queries. }
    \label{tab:result_r2med}
    \footnotesize 
    \setlength{\tabcolsep}{6pt} 
    \resizebox{\textwidth}{!}{
    \begin{tabular}{lllccccccccc}
        \toprule
        \multirow{2}{*}{\textbf{Paradigm}} & \multirow{2}{*}{\textbf{Models}} & \multirow{2}{*}{\textbf{Avg.}} & \multicolumn{3}{c}{\textbf{Q\&A Reference}} & \multicolumn{3}{c}{\textbf{Clinical Evidence}} & \multicolumn{2}{c}{\textbf{Clinical Case}} \\
        \cmidrule(lr){4-6} \cmidrule(lr){7-9} \cmidrule(lr){10-11}
        & & & \textbf{Bio.} & \textbf{Bioin.} & \textbf{MedS.} & \textbf{MedE.} & \textbf{MedD.} & \textbf{PMCT.} & \textbf{PMCC.} & \textbf{IIYiC.} \\
        \midrule
        Retriever & Diver-Retriever-4B & 42.9 & 49.3 & 60.4 & 57.9 & 18.6 & 22.6 & 58.7 & 44.1 & 31.8 \\ 
        \midrule
        \multirow{3}{*}{Pointwise} & RankT5-3B & 28.4 & 39.8 & 46.3 & 31.7 & 21.6 & 19.9 & 31.4 & 17.8 & 18.4 \\
        & ERank-4B & 41.8 & 47.9 & 63.0 & 62.3 & 21.7 & 28.7 & 41.4 & 46.5 & 22.7 \\
        & ERank-32B & 49.3 & 54.0 & \underline{69.5} & 62.2 & 25.0 & 31.9 & 65.9 & \underline{51.2} & \textbf{35.0} \\
        \midrule
        \multirow{7}{*}{Listwise} 
        & FIRST-7B & 28.2 & 34.1 & 45.1 & 37.0 & 14.0 & 15.1 & 37.3 & 23.2 & 19.9 \\
        & RankZephyr-7B & 37.5 & 48.8 & 55.7 & 54.9 & 26.7 & 26.1 & 57.8 & 10.9 & 18.7 \\ 
        & Rank-R1-7B & 47.8 & 58.2 & 65.8 & 56.7 & \underline{29.9} & \underline{40.6} & 66.1 & 35.1 & 29.8 \\
        & Rank-R1-14B & 49.3 & \underline{59.0} & 63.2 & 61.9 & 27.7 & 40.4 & 66.3 & 43.2 & 32.7 \\
        & Rank-K-32B & 49.5 & 58.9 & 67.8 & 58.9 & \textbf{30.3} & 40.5 & 65.3 & 42.5 & 31.5 \\
        & ReasonRank-7B & 42.8 & 48.9 & 61.6 & 61.7 & 20.5 & 32.4 & 59.1 & 36.9 & 21.5 \\
        & ReasonRank-32B & 50.2 & 53.9 & 68.9 & \textbf{68.4} & 28.2 & 40.3 & 67.4 & 45.1 & 29.1 \\ 
        \midrule
        & Qwen2.5-32B-Instruct & 48.4 & 53.0 & 65.3 & 64.8 & 24.2 & 31.7 & 70.3 & 46.1 & 31.5 \\
        & Qwen3-32B (No think) & 50.6 & 53.8 & 66.1 & \underline{67.9} & 28.7 & 40.1 & \underline{70.9} & 44.5 & 33.1 \\
        & Qwen3-32B (Think) & 50.7 & 55.2 & 66.9 & 64.9 & 26.9 & 39.9 & 68.3 & 48.5 & \underline{34.6} \\
        & DeepSeek-V3.1-671B   & \underline{52.0} & 54.1 & \textbf{70.0} & \underline{67.9} & \underline{29.9} & \textbf{41.5} & \textbf{71.1} & 50.1 & 31.8 \\
        \rowcolor[gray]{0.94}
        & GroupRank-7B & 47.8 & 56.7 & 65.4 & 64.2 & 27.3 & 30.5 & 64.6 & 43.2 & 30.9 \\
        \rowcolor[gray]{0.94}
        \multirow{-6}{*}{Groupwise} 
        & GroupRank-32B & \textbf{52.3} & \textbf{59.5} & \underline{69.7} & 66.3 & 28.8 & \textbf{41.5} & 67.6 & \textbf{52.2} & 32.7 \\
        \bottomrule
    \end{tabular}}
\end{table*}

\paragraph{Implementation} 
We follow a retrieval-and-reranking framework, employing a strong retriever DIVER-Retriever-4B~\citep{DIVER} to fetch the top-100 passages for subsequent ranking.
For the BRIGHT benchmark, we follow prior studies~\citep{ReasonRank,ERank} by using GPT4-rewritten queries for retrieval and original queries for reranking, while R2MED and BEIR utilize original queries for both stages. The reranker is built upon Qwen2.5-7B-Instruct and Qwen2.5-32B-Instruct \cite{qwen25} backbones, with hyperparameters set to $\gamma=0.2$ and $\epsilon=0.1$ in~\cref{eq:total_reward}. During inference, GroupRank uses group size $c=20$, while ReasonRank uses window 20 and step 10. Performance is measured by NDCG@10 (see \cref{sec:app-exp-setting}).

\subsection{Main Results}
\subsubsection{Results on Reasoning Benchmark}
The experimental results on the BRIGHT and R2MED benchmarks are reported in~\cref{tab:result_bright} and \cref{tab:result_r2med}. GroupRank consistently outperforms all baseline models, establishing new SOTA results for reasoning-intensive retrieval tasks.

\textbf{Compared with pointwise methods, groupwise ranking captures more discriminative relevance.}~Unlike pointwise methods that evaluate documents in isolation, our GroupRank perceives relative differences between multiple candidates simultaneously. On the BRIGHT benchmark, GroupRank-32B (38.0) significantly outperforms ERank-32B~(36.2), confirming that cross-document comparison is essential for capturing implicit relevance in complex scenarios. 


\textbf{Compared with listwise methods, the score-based paradigm facilitates hybrid ensembling.} While listwise models show promise, pointwise methods like ERank-32B remain competitive yet more efficient via parallelism. Specifically, ERank-32B outperforms ReasonRank-32B on BRIGHT and achieves comparable results on R2MED with lower latency. Unlike listwise models that output discrete sequences, groupwise and pointwise methods provide continuous scores. These scores enable weighted fusion with first-stage retrieval scores, allowing GroupRank to achieve superior final accuracy.

\textbf{High-quality data synthesis and specialized RL further unlock the potential of GroupRank.} Under the same setting, GroupRank achieves higher performance than both ERank and ReasonRank on both BRIGHT and R2Med. Notably, GroupRank-7B (34.3) outperforms all models with fewer than 32B parameters and even outperforms certain 32B-scale models (\textit{e.g.}, Rank-K) on the BRIGHT benchmark, highlighting the superior data efficiency and reasoning capabilities stimulated by our group-ranking reward.

\begin{table}[t]
  \centering
  \small
  \caption{Ablation study on BRIGHT. $\Delta$ indicates the NDCG@10 drop compared to the full model.}
  \label{tab:bright_ablation}
  \begin{tabular}{llc}
    \toprule
    Variant & NDCG@10 & $\Delta$ \\
    \midrule
    Diver-Retriever-4B & 32.9 & - \\
    \midrule
    \textbf{GroupRank} & \textbf{38.0} & - \\
    \quad w/o training (original model) & 32.6 & $\downarrow$ 5.4 \\
    \quad w/o SFT (only RL) & 31.7 & $\downarrow$ 6.3 \\
    \quad w/o RL (only SFT) & 33.4 & $\downarrow$ 4.6 \\
    \quad w/o Ranking-Utility Reward & 35.8 & $\downarrow$ 2.2 \\
    \quad w/o Group-Alignment Reward & 35.6 & $\downarrow$ 2.4 \\
    \bottomrule
  \end{tabular}
\end{table}

\subsubsection{Generalization on Semantic Benchmark} 
To assess the generalization of GroupRank on classic semantic-intensive tasks, we conduct additional experiments on the BEIR benchmark. As shown in \cref{tab:result_beir}, GroupRank-32B  outperforms strong listwise baselines and stands as one of the few models to deliver positive gains over the initial retriever. This confirms its effectiveness extends beyond reasoning-intensive scenarios to conventional semantic retrieval. Detailed analysis is provided in \cref{app:beir_results}.

\subsection{Ablation Study} \label{sec:ablation}
The ablation results in \cref{tab:bright_ablation} indicate that each module in GroupRank is essential for superior performance. SFT serves as the critical foundation, as its removal (leaving only RL) causes the largest performance drop of 6.3 points. This confirms that SFT is necessary to bootstrap the model into a favorable policy region for stable learning. While SFT provides the basic ranking capability, the RL stage further boosts NDCG@10 by 4.6 points through direct metric alignment. Within the reinforcement learning framework, the Group-Alignment Reward proves more impactful than the Ranking-Utility Reward. Removing the Group-Alignment Reward leads to a 2.4 point decline, indicating that direct optimization for document order is more vital for effective reranking than score calibration alone.

\begin{figure}[t] 
    \centering
    \includegraphics[width=\columnwidth]{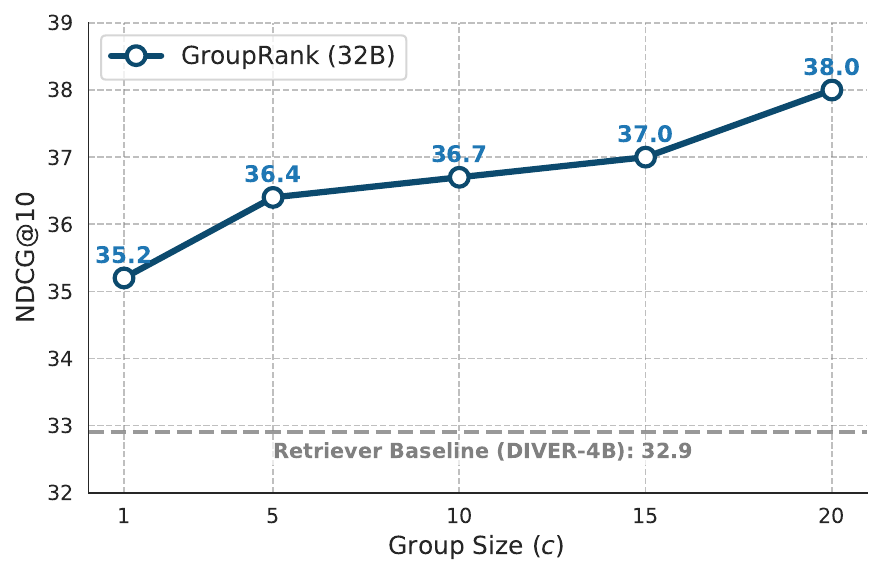}
    \caption{Effect of group size $c$ on NDCG@10 performance. The red dashed line represents the retriever baseline of 32.9. GroupRank reranks the top-100 passages retrieved by DIVER using GPT4-rewritten queries.}
    \label{fig:group_size_analysis}
\end{figure}

\subsection{More Analysis}
\paragraph{Effect of Group Size}
This section investigates the sensitivity of GroupRank to the hyperparameter group size $c$. As illustrated in  \cref{fig:group_size_analysis}, GroupRank consistently outperforms the retriever baseline of 32.9 across all configurations. Even with a minimum group size of $c=1$, the model achieves a substantial improvement with an NDCG@10 of 35.2. Performance scales positively as $c$ increases, reaching a peak of 38.0 at $c=20$. This steady upward trend demonstrates that larger group sizes facilitate more comprehensive cross-document interactions, which helps the model better identify subtle relevance differences among multiple candidates. The results indicate that the groupwise paradigm is highly effective at capturing fine-grained relevance and that its performance can be further enhanced by increasing the group size.

\begin{figure}[t]
    \centering
    \includegraphics[width=\columnwidth]{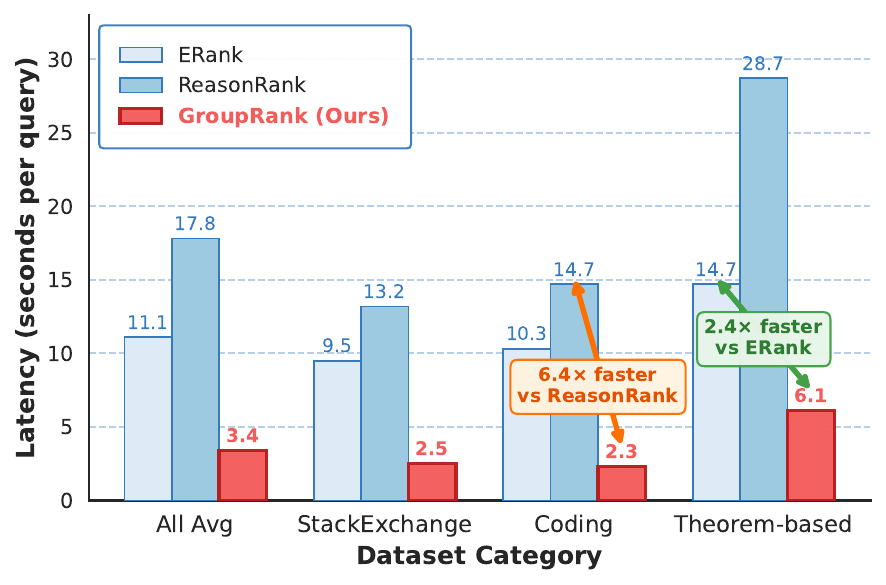}
    \caption{Average inference latency (seconds per query) on the BRIGHT benchmark. All methods are evaluated using 32B models on four H800-80G GPUs. }
    \label{fig:ranking_latency}
\end{figure}

\begin{table*}[t]
\centering
\caption{Impact of Group Formation on the BRIGHT benchmark (NDCG@10).}
\label{tab:group_formation_bright}
\footnotesize
\setlength{\tabcolsep}{3pt}
\resizebox{\textwidth}{!}{
\begin{tabular}{llcccccccccccccc}
\toprule
Paradigm & Method & Avg. & Bio. & Ear. & Eco. & Psy. & Rob. & Sta. & Sus. & Lee. & Pon. & AoP. & ThQ. & ThT. \\
\midrule
Retriever & Diver-Retriever-4B & 32.9 & 52.5 & 53.6 & 33.8 & 45.2 & 28.4 & 30.4 & 35.0 & 13.0 & 14.7 & 9.8 & 42.5 & 36.3 \\
\midrule
Groupwise & GroupRank-32B (Sorted) & 36.2 & 55.9 & 52.6 & 38.6 & 49.0 & 37.8 & 34.5 & 43.2 & 13.2 & 15.7 & 10.5 & 37.9 & 46.0 \\
Groupwise & GroupRank-32B (Random) & 38.0 & 59.0 & 57.5 & 39.2 & 50.0 & 39.1 & 39.0 & 42.7 & 14.3 & 14.9 & 12.6 & 39.0 & 48.8 \\
\bottomrule
\end{tabular}
}
\end{table*}

\begin{table*}[t]
\centering
\caption{Impact of Group Formation on the R2MED benchmark (NDCG@10).}
\label{tab:group_formation_r2med}
\footnotesize
\setlength{\tabcolsep}{6pt}
\resizebox{\textwidth}{!}{
\begin{tabular}{llccccccccc}
\toprule
Paradigm & Method & Avg. & Bio. & Bion. & MedS. & MedE. & MedD. & PMCT. & PMCC. & IIYiC. \\
\midrule
Retriever & Diver-Retriever-4B & 42.9 & 49.3 & 60.4 & 57.9 & 18.6 & 22.6 & 58.7 & 44.1 & 31.8 \\
\midrule
Groupwise & GroupRank-32B (Sorted) & 49.8 & 56.0 & 68.9 & 64.6 & 28.8 & 39.0 & 64.7 & 46.8 & 29.9 \\
Groupwise & GroupRank-32B (Random) & 52.3 & 59.5 & 69.7 & 66.3 & 28.8 & 41.5 & 67.6 & 52.2 & 32.7 \\
\bottomrule
\end{tabular}
}
\end{table*}
\paragraph{Impact of Group Formation}
\label{sec:group_formation}
Since input ordering determines group formation, it directly affects how documents are compared. We compare our default random grouping with a sorted strategy that partitions candidates by their initial rank. Unlike sorted grouping, which clusters highly similar documents together, random grouping shuffles the list to provide more diverse comparative contexts.

As shown in \cref{tab:group_formation_bright,tab:group_formation_r2med}, random grouping yields notably superior results. We attribute this to the fact that comparing nearly identical documents forces the model to focus on indistinguishable nuances, potentially introducing scoring bias or noise. Shuffling instead allows each document to be calibrated against a broader range of candidates, leading to more objective and stable relevance estimations. These findings confirm the significance of input ordering and establish random grouping as the more effective strategy.

\paragraph{Inference Efficiency Analysis}
The inference latency of GroupRank is evaluated against pointwise and listwise baselines on the BRIGHT benchmark. As shown in~\cref{fig:ranking_latency}, GroupRank achieves the lowest latency, providing a \textbf{2.4$\times$ to 6.4$\times$} speedup over competitive baselines. Specifically, GroupRank processes a query in 3.4 seconds on average, while the pointwise ERank takes 11.1 seconds. On complex theorem-based tasks, GroupRank completes in 6.1 seconds, whereas ReasonRank takes 28.7 seconds. 
These gains stem from the groupwise paradigm's structural advantages: it minimizes redundant model calls compared to pointwise scoring and, unlike sequential listwise sliding windows, enables parallel processing across groups. GroupRank thus offers a scalable solution for latency-sensitive applications, with further theoretical analysis in~\cref{sec:theoretical_efficiency}.

\begin{figure}[t]
    \centering
    \includegraphics[width=1.0\linewidth]{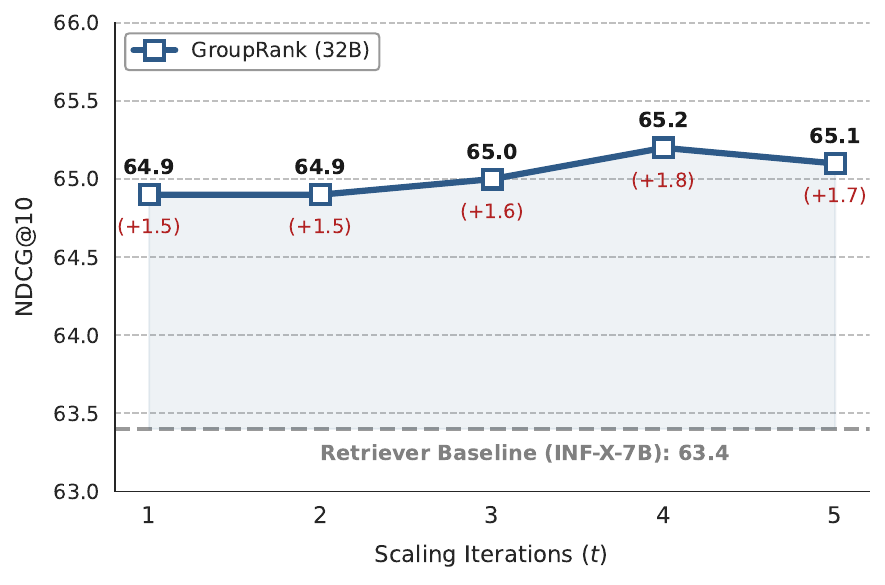}
    \caption{Enhanced Performance of GroupRank with advanced retrieval strategies and test-time scaling.}
    \label{fig:scaling_results}
\end{figure}

\subsection{Further Enhancement} \label{sec:exp-further_enhancement}
We further explore GroupRank's potential performance by integrating it with query rewriting and retrieval frameworks. 
Using INF-rewritten queries and the INF-X-Retriever~\citep{inf-x-retriever-2025}, GroupRank improves the initial baseline from 63.4 to 64.9, as shown in \cref{fig:scaling_results}. To further enhance performance, we apply test-time scaling~\citep{retro_ranker}, which shuffles documents into different group combinations across multiple iterations and averages their resulting scores. This process ensures each document is evaluated within various comparative contexts, effectively reducing the bias of random partitioning. Performance consistently increases with iterations, reaching a \textbf{new SOTA of 65.2 NDCG@10}. While gains saturate after four iterations, these results demonstrate that GroupRank successfully leverages scaling to maximize ranking quality within advanced retrieval pipelines.

\section{Conclusion}
This paper proposes GroupRank, a groupwise reranking framework balancing accuracy and efficiency for LLM-based passage retrieval. By processing document groups in parallel, GroupRank captures essential cross-document interactions while alleviating the latency issues. The training process utilizes an answer-free synthesis pipeline for SFT and RL, where a specialized group-ranking reward optimizes document ranking and score calibration. Experimental results show that GroupRank achieves SOTA performance with high efficiency, providing a robust and practical solution for reasoning-intensive scenarios.

\section*{Limitations and Future Work}
GroupRank achieves promising ranking performance, but still faces some limitations. First, its random document grouping may cluster highly relevant candidates together, which forces the model to differentiate among similarly relevant documents. This can lead to exaggerated score gaps within groups and introduce ranking bias, especially in single-pass inference. Although the test-time scaling strategy mitigates this issue by averaging scores from multiple group allocations, bias remains for real-time inference settings with limited iterations.
Second, we have not yet trained GroupRank on non-reasoning models to further reduce inference latency. For fair comparison with previous baselines, our current implementation uses reasoning-based models such as Qwen2.5-32B as backbone. Reasoning enables the model to better distinguish subtle differences between documents, enhancing ranking accuracy but also increasing latency. Future work will focus on adapting the groupwise paradigm to smaller, non-reasoning models to achieve lower latency while maintaining competitive ranking performance.

\section*{Acknowledgments}
This work is supported by the National Natural Science Foundation of China (62472461), the Guangdong Basic and Applied Basic Research Foundation (2025A1515010129), and Ant Group Research Intern Program. 

\bibliography{reference}

\clearpage
\appendix
\section{Dataset and Training Details} \label{sec:app-Data Synthesis}

\subsection{Dataset Attributes and Composition}
In this section, we provide a detailed introduction to the statistical characteristics of the training data. The query distribution is presented as~\cref{tab:statistics}.

\begin{table}[h]
\centering
\caption{Statistics of the training queries (totally 13.6k).}
\label{tab:statistics}
\resizebox{\columnwidth}{!}{
\setlength{\tabcolsep}{12pt}
\begin{tabular}{llr}
\toprule
\textbf{Category} & \textbf{Dataset Name} & \textbf{Count} \\ 
\midrule
\multirow{7}{*}{Complex QA} 
 & Biology            & 1,700 \\
 & Earth Science      & 566   \\
 & Economics          & 787   \\
 & Robotics           & 451   \\
 & Sustainable Living & 147   \\
 & Stackoverflow      & 1,741 \\
 & Medical            & 1,000 \\ 
\cmidrule(lr){1-3}
\multirow{1}{*}{Coding} & xCodeEval          & 700   \\
\cmidrule(lr){1-3}
\multirow{2}{*}{Math} 
 & Math-QA            & 1,726 \\
 & Math-Theorem       & 1,673 \\ 
\cmidrule(lr){1-3}
Web Search & MS MARCO           & 3,093 \\ 
\bottomrule
\end{tabular}
}
\end{table}

To ensure the GroupRank model possesses the flexibility to handle candidate lists of arbitrary lengths during inference, we constructed our training datasets with a diverse range of group sizes ($c$). As detailed in~\cref{tab:group_size_dist}, the group size distribution covers a spectrum from 5 to 20 for both the Supervised Fine-Tuning (SFT) and Reinforcement Learning (RL) phases. While the distribution is strategically weighted towards standard sizes of $c=10$ and $c=20$ to facilitate effective training alignment—accounting for approximately 93.5\% of the SFT data (9,321 samples for $c=10$, 8,711 for $c=20$) and 79.9\% of the RL data—we deliberately incorporated samples across other sizes (\textit{e.g.}, sizes 5-9 and 11-19). This strategy prevents the model from overfitting to specific input dimensions, thereby enhancing its robustness and generalization capabilities when processing variable-sized document groups in real-world retrieval scenarios.

\subsection{Effectiveness of Dual-Teacher Fusion}  \label{sec:app-Dual-Teacher-Fusion}
The effectiveness of dual-teacher framework is empirically validated on the BRIGHT benchmark. As shown in \Cref{tab:teacher_fuse_experiment}, the fusion of pointwise and listwise signals yields superior performance compared to either approach individually. This indicates the complementary nature of the two paradigms: pointwise scoring provides fine-grained document-level relevance assessment, while listwise ranking captures inter-document relationships from a global perspective. The consistent performance gains across diverse domains confirm that the fusion strategy generates higher-quality pseudo-labels for training.

\begin{table}[!t]
    \centering
    \small
    \caption{Distribution of group sizes in SFT and RL training datasets.}
    \label{tab:group_size_dist}
    \begin{tabular}{l|cc}
    \toprule
    \textbf{Group Size} & \textbf{SFT Count} & \textbf{RL Count} \\
    \midrule
    \textbf{5}   & 102   & 85   \\
    \textbf{6}   & 88    & 82   \\
    \textbf{7}   & 72    & 75   \\
    \textbf{8}   & 97    & 77   \\
    \textbf{9}   & 80    & 81   \\
    \textbf{10}  & 9,321 & 2,389 \\
    \textbf{11}  & 84    & 94   \\
    \textbf{12}  & 86    & 79   \\
    \textbf{13}  & 87    & 82   \\
    \textbf{14}  & 89    & 80   \\
    \textbf{15}  & 97    & 76   \\
    \textbf{16}  & 101   & 77   \\
    \textbf{17}  & 86    & 73   \\
    \textbf{18}  & 82    & 89   \\
    \textbf{19}  & 79    & 75   \\
    \textbf{20}  & 8,711 & 2,393 \\
    \midrule
    \textbf{Total} & 19,262 & 5,907 \\
    \bottomrule
    \end{tabular}
\end{table}

\begin{table}[!t]
\centering
\small
\caption{Hyperparameters and settings for SFT.}
\label{tab:sft_params}
\begin{tabular}{l|c}
\toprule
\textbf{Hyperparameter} & \textbf{Value}\\
\midrule
Base Model & Qwen2.5-7B/32B-Instruct \\
Training Framework & MS-Swift \\
Training Strategy & Full \\
Optimization Strategy & DeepSpeed ZeRO-3 \\
Attention Implementation & FlashAttention-2 \\
\midrule
Precision  & bfloat16 \\
Epochs & 5\\
Learning Rate & $3 \times 10^{-5}$ \\
Max Sequence Length & 32,768\\
Per-Device Batch Size & 2\\
Gradient Accumulation & 4 \\
Sequence Packing & True \\
\bottomrule
\end{tabular}
\end{table}

\begin{table*}[t]
\centering
\caption{Performance comparison of pseudo-labeling strategies on BRIGHT benchmark.}
\label{tab:teacher_fuse_experiment}
\small
\setlength{\tabcolsep}{3.5pt}
\resizebox{\textwidth}{!}{
\begin{tabular}{llc ccccccc cc ccc}
\toprule
\multirow{2}{*}{\textbf{Paradigm}} & \multirow{2}{*}{\textbf{Method}} & \multirow{2}{*}{\textbf{Avg.}} & \multicolumn{7}{c}{\textbf{StackExchange}} & \multicolumn{2}{c}{\textbf{Coding}} & \multicolumn{3}{c}{\textbf{Theorem-based}} \\
\cmidrule(lr){4-10} \cmidrule(lr){11-12} \cmidrule(lr){13-15}
& & & \textbf{Bio.} & \textbf{Ear.} & \textbf{Eco.} & \textbf{Psy.} & \textbf{Rob.} & \textbf{Sta.} & \textbf{Sus.} & \textbf{Lee.} & \textbf{Pon.} & \textbf{Aop.} & \textbf{ThQ.} & \textbf{ThT.} \\
\midrule
Pointwise & Qwen3-235B-Instruct & 36.3 & 57.3 & 55.1 & 36.4 & 51.5 & 35.9 & 34.4 & 40.8 & 11.5 & 21.7 & 9.1 & 37.6 & 44.7 \\
Listwise  & Gemini-2.5-Flash & 38.6 & 56.6 & 56.4 & 38.5 & 53.1 & 40.1 & 35.7 & 47.3 & 15.5 & 25.3 & 10.6 & 39.5 & 45.2 \\
\textbf{Fuse} & \textbf{Pointwise+Listwise} & \textbf{39.8} & \textbf{59.9} & \textbf{59.0} & \textbf{38.6} & \textbf{55.3} & \textbf{41.4} & \textbf{35.2} & \textbf{48.0} & \textbf{13.5} & \textbf{27.9} & \textbf{11.1} & \textbf{40.2} & \textbf{47.9} \\
\bottomrule
\end{tabular}
}
\end{table*}

\begin{algorithm}[!t]
\caption{Pseudo-Label Synthesis}
\label{alg:golden_list_synthesis}
\begin{algorithmic}[1]
\renewcommand{\algorithmiccomment}[1]{\hfill $\rhd$ \textit{\footnotesize #1}}

\Require Query $q$, documents $\mathcal{D} = \{d_1, \dots, d_N\}$ ($N=50$)
\Ensure Training samples $\mathcal{X}_{\text{train}}$

\Statex \textbf{// Step 1: Dual-Teacher ranking}
\State $\mathbf{s}_p \gets [T_p(q, d_i)]_{i=1}^N$ \Comment{pointwise scores}
\State $L_p \gets \text{argsort}(\mathbf{s}_p, \text{desc})$ \Comment{pointwise ranks}
\State $L_l \gets T_l(q, \mathcal{D})$ \Comment{listwise ranks}

\Statex \textbf{// Step 2: Ranking fusion}
\For{$d \in \mathcal{D}$}
    \State $S_{\text{label}}(d) \gets -\beta \log L_p(d)-(1-\beta) \log L_l(d)$
\EndFor
\State $L_{label} \gets \text{argsort}([S_{\text{label}}(d_i)]_{i=1}^N,\,\text{desc})$ 

\Statex \textbf{// Step 3: Multi-scale sampling}
\State $\mathcal{X}_{\text{train}} \gets \emptyset$
\For{$G \in \{5, \dots, 20\}$}
    \State $\mathcal{D}_G \gets \text{sample}(L_{label}, G)$ \Comment{interval sampling}
    \State $\mathbf{y}_G \gets \text{project}(L_{label}, \mathcal{D}_G)$ \Comment{sub-list ranks}
    \State $\mathcal{X}_{\text{train}} \gets \mathcal{X}_{\text{train}} \cup \{(q, \mathcal{D}_G, \mathbf{y}_G)\}$
\EndFor

\State \Return $\mathcal{X}_{\text{train}}$
\end{algorithmic}
\end{algorithm}

\subsection{Multi-Scale Data Generation Algorithm}
\cref{alg:golden_list_synthesis} describes pseudo-label synthesis strategy for generating high-quality training data across multiple scales. By generating a single global ranking for the full document set and subsequently sampling sub-lists of varying sizes, we achieve approximately 10× cost reduction compared to invoking teacher models for each individual group. This strategy maintains data diversity while significantly optimizing computational resources.

\subsection{Training Details} \label{sec:app-exp-setting}
We perform \textbf{SFT} on Qwen2.5-Instruct models using the MS-Swift framework. Specifically, we apply full-parameter fine-tuning to the 7B model, while employing LoRA for the larger 32B model to ensure training efficiency. This stage is conducted on a distributed cluster of 4 nodes, each equipped with 8 NVIDIA H200 GPUs. Specific hyperparameters are provided in~\cref{tab:sft_params}.
For the \textbf{RL} training stage, we scale the infrastructure to 8 nodes (also H200 GPUs). We utilize FSDP (Fully Sharded Data Parallel) with parameter and optimizer offloading to mitigate memory constraints. To accelerate the generation phase, we integrate vLLM as the rollout engine with a tensor parallel size of 8. The learning rate is set to $1\text{e-}5$, using a low-variance KL estimator ($\phi=0.001$) for stability. Detailed hyperparameters are listed in~\cref{tab:grpo_params}.

\begin{figure}[!t] 
    \centering
    \includegraphics[width=\columnwidth]{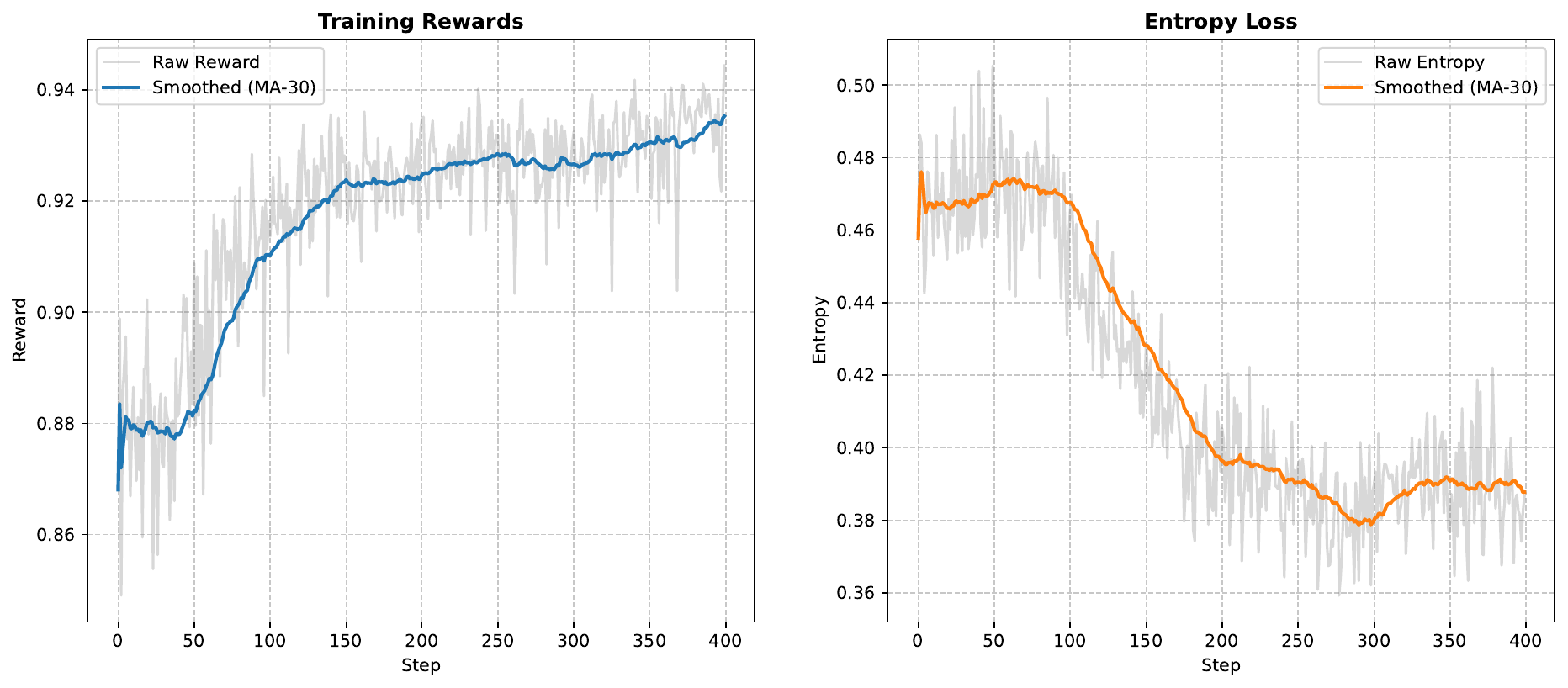}
    \caption{Training curves of GroupRank. The left panel shows average reward per step, demonstrating a steady increase over time. The right panel displays the entropy loss, which decreases as the agent’s policy converges. The solid lines represent the smoothed moving average, while the shaded regions indicate the raw fluctuation.}
    \label{fig:reward}
\end{figure}

As shown in \cref{fig:reward}, the reward curve demonstrates a consistent upward trend, indicating that the agent effectively learns the optimal policy over time. In the initial phase (\textit{e.g.}, first 1,000 steps), the rewards fluctuate significantly due to exploration. 
However, as training progresses, the curve stabilizes and converges to a high value, demonstrating the stability and convergence of GroupRank.

\begin{table}[t]
\centering
\small
\caption{Hyperparameters and implementation details for the GroupRank training using GRPO.}
\label{tab:grpo_params}
\begin{tabular}{l|c}
\toprule
\textbf{Hyperparameter} & \textbf{Value} \\
\midrule
\multicolumn{2}{l}{\textit{Optimization \& LoRA Settings}} \\
\midrule
Algorithm & GRPO \\
Fine-tuning Method & LoRA (Rank=32, $\alpha$=32) \\
Target Modules & All Linear Layers \\
Learning Rate & $1 \times 10^{-5}$ \\
Total Epochs & 5 \\
Global Train Batch Size & 64 \\
PPO Mini-Batch Size & 64 \\
\midrule
\multicolumn{2}{l}{\textit{RL \& Generation Details}} \\
\midrule
Group Size (Rollout $N$) & 8 \\
KL Coefficient ($\phi$) & 0.001 \\
KL Type & Low-Variance KL \\
Generation Temperature & 1.0 (Default) \\
Max Prompt Length & 24,000 \\
Max Response Length & 8,000 \\
\midrule
\multicolumn{2}{l}{\textit{Infrastructure (Verl \& vLLM)}} \\
\midrule
Rollout Engine & vLLM \\
Tensor Parallel Size & 8 \\
FSDP Size & 8 \\
Offloading strategy & Parameter \& Optimizer Offload \\
Nodes & 8 \\
GPUs per Node & 8 \\
\bottomrule
\end{tabular}
\end{table}

\subsection{Prompts}
We design three distinct prompts to facilitate our proposed pipeline. For data synthesis, we employ a Listwise Labeling Prompt~\ref{prompt:listwise_labeling} and a Pointwise Labeling Prompt~\ref{prompt:pointwise_labeling}, whose outputs are fused to construct a robust hybrid training dataset. Crucially, we introduce a Groupwise Ranking Prompt~\ref{prompt:groupwise}, which serves as the consistent instruction for both supervising the model during training and executing the ranking task during inference.

\section{Evaluation Setup} \label{sec:app-experiments} 
\subsection{Benchmarks} 
We conduct comprehensive evaluations on three distinct benchmarks to assess different facets of our model: BRIGHT for reasoning-intensive retrieval, R2MED for domain-specific medical expertise, and BEIR for general zero-shot retrieval capabilities. Details of each benchmark are described below:

\textbf{BRIGHT} is a benchmark specifically constructed for reasoning-intensive retrieval. Unlike traditional datasets that rely on semantic overlap or keyword matching, BRIGHT collects real-world queries from domains requiring deep cognitive processing, such as software engineering (\textit{e.g.}, LeetCode), mathematics, and logic puzzles. A defining characteristic of this benchmark is the substantial lexical gap between queries and relevant documents; successful retrieval depends on logical deduction and multi-step reasoning rather than surface-level textual similarity.
 
\textbf{R2MED} is a specialized benchmark designed to evaluate information retrieval and reranking systems within the biomedical and clinical domains. The dataset aggregates tasks that involve complex medical terminology and specific professional contexts. It serves as a standard for assessing a system's proficiency in handling domain-specific knowledge, requiring precise understanding of medical entities and the ability to distinguish relevant clinical evidence from noise.

\textbf{BEIR} is a heterogeneous benchmark developed for the evaluation of zero-shot information retrieval across diverse distributions. 

\subsection{Baselines}
We benchmark against a broad spectrum of reranking paradigms, ranging from fine-tuned encoder-decoder models (RankT5) and general LLM-based listwise approaches (RankZephyr, ERank) to the latest reasoning-enhanced frameworks (Rank-R1, Rank-K, ReasonRank).

\textbf{RankT5}: An encoder-decoder ranking model that fine-tunes T5 using listwise or pairwise ranking losses to directly output relevance scores.

\textbf{ERank}: A hybrid framework that fuses Supervised Fine-Tuning (SFT) with Reinforcement Learning (RL) to enhance both the effectiveness and efficiency of text reranking.

\textbf{FIRST}: An efficient listwise reranking method that improves decoding speed by optimizing logit generation within a single window, though it still relies on sequential sliding windows to handle long candidate lists.

\textbf{RankZephyr}: An effective zero-shot listwise reranking method that distills the ranking capabilities of GPT-4 into an open-source 7B model (Zephyr) via instruction tuning.

\textbf{Rank-R1}: A reasoning-enhanced reranker that utilizes Reinforcement Learning with outcome-driven rewards to stimulate reasoning capabilities without requiring explicit reasoning supervision. 

\textbf{Rank-K}: A listwise reranker designed for "test-time reasoning" that distills reasoning traces from large reasoning models (like DeepSeek-R1 or QwQ) to handle complex queries.

\textbf{ReasonRank}: A reasoning-intensive reranking framework that employs a two-stage training strategy (SFT on synthetic reasoning data followed by RL) to empower the model with strong logical reasoning abilities.

\subsection{Evaluation Metrics}
To evaluate the ranking quality, we employ the Normalized Discounted Cumulative Gain at rank 10 (\textbf{NDCG@10}). Unlike simple recall metrics, NDCG accounts for the position of relevant documents by assigning higher scores to hits at the top of the list. It is calculated as the Discounted Cumulative Gain (DCG) divided by the Ideal DCG (IDCG), which represents the score of a perfect ranking. The metric is defined as:
\begin{equation}
\text{NDCG}@K = \frac{\text{DCG}@K}{\text{IDCG}@K}
\end{equation}
where $\text{DCG}@K$ is computed as:
\begin{equation}
\text{DCG}@K = \sum_{i=1}^{K} \frac{rel_i}{\log_2(i+1)}
\end{equation}

\begin{table*}[t]
  \centering
  \caption{NDCG@10 on the BEIR benchmark. Models rerank the top-100 passages retrieved with original queries.}
  \label{tab:result_beir}
  \footnotesize 
  \setlength{\tabcolsep}{5pt} 
  \resizebox{0.8\textwidth}{!}{
  \begin{tabular}{lllccccc}
    \toprule
    \textbf{Paradigm} & \textbf{Models} & \textbf{Avg.} & \textbf{Arguana} & \textbf{Dbpedia} & \textbf{Nfcorpus} & \textbf{NQ} & \textbf{Scidocs} \\
    \midrule
    Retriever & Diver-Retriever-4B & \underline{52.9} & \textbf{34.2} & 71.4 & 44.8 & 89.7 & \underline{24.3} \\ 
    \midrule
    \multirow{3}{*}{Pointwise} & RankT5-3B & 43.0 & 11.2 & 58.9 & 35.3 & 91.8 & 17.7 \\
    & ERank-4B & 42.4 & 26.5 & 64.4 & 3.3 & 95.6 & 22.5 \\ 
    & ERank-32B & 42.5 & 26.7 & 65.9 & 3.1 & \underline{95.8} & 21.0 \\ 
    \midrule
    \multirow{6}{*}{Listwise} 
    & RankZephyr-7B & 44.2 & 12.1 & 59.8 & 37.1 & 93.1 & 19.0 \\
    & Rank-R1-7B & 44.7 & 14.3 & 61.5 & 36.6 & 92.0 & 19.0 \\
    & Rank-R1-14B & 50.1 & 17.1 & 78.2 & 41.0 & 92.4 & 21.7 \\
    & Rank-K-32B & 52.1 & 22.2 & 78.4 & \underline{45.3} & 93.3 & 21.4 \\
    & ReasonRank-7B & 48.6 & 19.2 & 69.8 & 40.3 & 91.5 & 22.2 \\
    & ReasonRank-32B & 52.1 & 18.5 & \underline{79.0} & 44.4 & 94.7 & 24.0 \\ 
    \midrule
    \rowcolor[gray]{0.94}
    & GroupRank-7B & 46.5 & \underline{30.2} & 71.2 & 40.2 & 76.2 & 14.5 \\
    \rowcolor[gray]{0.94}
    \multirow{-2}{*}{Groupwise} & GroupRank-32B & \textbf{55.1} & 25.0 & \textbf{82.0} & \textbf{46.6} & \textbf{96.2} & \textbf{25.7} \\
    \bottomrule
  \end{tabular}}
\end{table*}

\begin{table*}[t]
    \centering
    \caption{NDCG@10 results of ablation study on the BRIGHT benchmark}
    \label{tab:result_bright_ablations}
    \footnotesize 
    \setlength{\tabcolsep}{4.2pt} 
    \resizebox{\textwidth}{!}{
    \begin{tabular}{llcccccccccccc}
        \toprule
        \textbf{Models} & \textbf{Avg.} & \textbf{Bio.} & \textbf{Earth.} & \textbf{Econ.} & \textbf{Psy.} & \textbf{Rob.} & \textbf{Stack.} & \textbf{Sus.} & \textbf{Leet.} & \textbf{Pony} & \textbf{AoPS} & \textbf{TheoQ.} & \textbf{TheoT.} \\
        \midrule
        GroupRank (32B) & \textbf{38.0} & \textbf{59.0}	&\textbf{57.5}	& \textbf{39.2}	& \textbf{50.0}	& \textbf{39.1}	&\textbf{39.0}	&\textbf{42.7}	&\textbf{14.3}	&14.9	&\textbf{12.6}	& \textbf{39.0}	&\textbf{48.8} \\
        \midrule
        w/o training & 32.6 &51.6	&46.4	&31.6	&43.3	&26.8	&32.5	&40.7	&14.0	&18.0	&8.8	&33.4	&44.2 \\
        w/o SFT (only RL) &31.7 & 52.4  & 44.9 & 31.8 & 43.2     & 33.3 & 33.1 & 41.5 & 13.3 & \textbf{20.4} & 8.3 & 36.0   & 43.6 \\
        w/o RL (only SFT)  &33.4 &53.2	&51.4	&37.3	&47.0	&35.7	&35.8	&42.3	&13.3	&11.4	&12.3	&38.2	&43.2 \\
        w/o Ranking-Utility Reward &35.8 &53.2  &54.2      &38.8  & 50.0     &33.1 &35.7 &42.6 &13.3   &12.1   &12.0   &39.0   &45.7 \\
        w/o Group-Alignment Reward  &35.6 &53.4          &52.2      &38.3   &49.0  &33.2   & 36.5    & 41.6     &13.0   &12.0  & 11.6  & 38.3   &48.0   \\
        \bottomrule
    \end{tabular}}
\end{table*}

\section{More Experiments}
\subsection{Results on Semantic Benchmark}
\label{app:beir_results}
To verify the generalization of GroupRank in traditional semantic retrieval scenarios, we evaluate its performance on representative subsets of the BEIR benchmark \citep{beir}. 
\cref{tab:result_beir} shows that GroupRank-32B achieves an average NDCG@10 of 55.1, outperforming strong listwise baselines like Rank-K-32B and ReasonRank-32B by a margin of 3.0 points. This success confirms that the effectiveness of GroupRank extends beyond reasoning-intensive scenarios to general semantic search, as exemplified by its leading scores on Dbpedia (82.0) and NQ (96.2).

Notably, the results reveal that \textbf{many competitive rerankers fail to surpass the initial retrieval score of 52.9}. For instance, the pointwise model ERank-32B drastically underperforms the retriever by over 10 points. This indicates that surpassing a strong retriever on semantic-dense datasets is difficult, as improper reranking can introduce noise. In contrast,\textbf{ GroupRank-32B is one of the few models to deliver consistent positive gains}. This resilience proves that the groupwise paradigm captures robust relevance patterns, establishing it as a universal solution for diverse retrieval applications.

\subsection{Details of Ablation Study}
\cref{tab:result_bright_ablations} shows the detailed ablation results on the subset of BRIGHT. All subsets' metrics demonstrate that each module in GroupRank is essential for better performance. About 9 subsets show that SFT serves as the critical foundation, as its removal causes the largest performance drop of 3-11.11 points. The RL stage is also very important because its removal causes 2-7 points in all subsets of BRIGHT. The group-alignment reward is relatively more important than the Ranking-Utility reward.

\begin{table*}[t]
\centering
\caption{Efficiency comparison of different reranking paradigms. $N$ is the number of documents, $k$ is the top-$k$ retrieved documents to rerank, and $c, r, s$ represent the group size, the number of reranking repeats, and the sliding window step size, respectively. Generate indicates whether it is a generative model, Parallel denotes its support for parallel processing, Cross-Document represents the capability for cross-document comparison, and TTS signifies whether it supports Test-time scaling.}
\label{tab:theoretical_efficiency}
\resizebox{0.8\textwidth}{!}{
\begin{tabular}{lccccc}
\toprule
\textbf{Method} & \textbf{Generate} & \textbf{Parallel} & \textbf{Cross-Document} & \textbf{TTS}& \textbf{Complexity} \\ 
\midrule
Pointwise (qlm)        &            & \checkmark & & \checkmark & $\mathcal{O}(N)$ \\
Pointwise (yes/no)     &            & \checkmark & & \checkmark & $\mathcal{O}(N)$ \\
Listwise (generation)  & \checkmark &            &\checkmark  &  & $\mathcal{O}(\frac{r \cdot N}{s})$ \\
Listwise (likelihood)  &            &            &\checkmark  &  & $\mathcal{O}(\frac{r \cdot N}{s})$ \\
Pairwise (all-pair)    & \checkmark & \checkmark &\checkmark  &  & $\mathcal{O}(N^2)$ \\
Pairwise (heapsort)    & \checkmark &            &\checkmark  &  & $\mathcal{O}(k \log_{2}{N})$ \\
Pairwise (bubblesort)  & \checkmark &            &\checkmark  &  & $\mathcal{O}(k \cdot N)$ \\
Setwise (heapsort)     & \checkmark &            &\checkmark  &  & $\mathcal{O}(k \log_{c}{N})$ \\
Setwise (bubblesort)   & \checkmark &            &\checkmark  &  & $\mathcal{O}(\frac{k \cdot N}{c-1})$ \\
\midrule
\textbf{Groupwise (ours)}     & \checkmark & \checkmark & \checkmark & \checkmark & $\mathcal{O}(\frac{N}{c})$ \\
\bottomrule
\end{tabular}
}
\end{table*}

\begin{table*}[t]
    \centering
    \caption{Performance comparison on the BRIGHT benchmark using different retrievers. Best results are \textbf{bolded}.}
    \label{tab:retriever_comparison_v2}
    \footnotesize 
    \setlength{\tabcolsep}{3.8pt} 
    \resizebox{\textwidth}{!}{
    \begin{tabular}{llc ccccccc cccc cc}
        \toprule
        \multirow{2}{*}{\textbf{Paradigm}} & \multirow{2}{*}{\textbf{Method}} & \multirow{2}{*}{\textbf{Avg.}} & \multicolumn{7}{c}{\textbf{StackExchange}} & \multicolumn{2}{c}{\textbf{Coding}} & \multicolumn{3}{c}{\textbf{Theorem-based}} \\
        \cmidrule(lr){4-10} \cmidrule(lr){11-12} \cmidrule(lr){13-15}
        & & & \textbf{Bio.} & \textbf{Earth.} & \textbf{Econ.} & \textbf{Psy.} & \textbf{Rob.} & \textbf{Stack.} & \textbf{Sus.} & \textbf{Leet.} & \textbf{Pony} & \textbf{AoPS} & \textbf{TheoQ.} & \textbf{TheoT.} \\
        
        \midrule
        \rowcolor[gray]{0.98} 
        \multicolumn{15}{c}{\rule{0pt}{2ex} \textit{Setting 1: Using BM25 Retriever} \rule{0pt}{2ex}} \\
        \midrule
        Retriever & BM25 & 27.0 & 53.6 & 54.1 & 24.3 & 38.7 & 18.9 & 27.7 & 26.3 & 19.3 & 17.6 & 3.9 & 19.2 & 20.8 \\
        Pointwise & ERank (32B) & 33.5 & 60.0 & 59.7 & 30.5 & 46.9 & 23.2 & 29.3 & 36.0 & 15.2 & \textbf{28.0} & 8.0 & 26.8 & 38.0 \\
        Listwise & ReasonRank (32B) & 35.5 & 54.3 & 47.6 & 33.1 & \textbf{50.8} & \textbf{32.8} & 34.8 & \textbf{43.4} & 24.7 & 24.1 & 9.1 & \textbf{31.4} & \textbf{40.3} \\
        Groupwise & GroupRank (32B) & \textbf{37.3} & \textbf{62.5} & \textbf{59.8} & \textbf{34.7} & 50.3 & 31.9 & \textbf{37.6} & 42.2 & \textbf{28.3} & 21.5 & \textbf{10.4} & 30.2 & 38.4 \\
        
        \midrule
        \rowcolor[gray]{0.98}
        \multicolumn{15}{c}{\rule{0pt}{2ex} \textit{Setting 2: Using ReasonIR-8B Retriever} \rule{0pt}{2ex}} \\
        \midrule
        Retriever & ReasonIR (8B) & 30.6 & 43.5 & 43.0 & 32.8 & 38.9 & 21.1 & 31.1 & 27.3 & \textbf{31.6} & 19.7 & 7.3 & 34.1 & 36.7 \\
        Pointwise & ERank (32B) & 36.6 & 54.7 & 51.2 & 35.4 & 44.3 & 24.8 & 37.2 & 38.2 & 29.7 & \textbf{31.5} & 9.7 & 37.1 & 45.6 \\
        Listwise & ReasonRank (32B) & 36.0 & 56.3 & 46.1 & 35.8 & \textbf{52.3} & \textbf{32.7} & 33.1 & 42.3 & 24.6 & 19.5 & 8.1 & 38.0 & 43.3 \\
        Groupwise & GroupRank (32B) & \textbf{38.2} & \textbf{58.5} & \textbf{51.9} & \textbf{38.1} & 50.6 & 29.3 & \textbf{40.3} & \textbf{45.3} & 28.2 & 20.0 & \textbf{10.5} & \textbf{39.8} & \textbf{45.8} \\

        \midrule
        \rowcolor[gray]{0.98}
        \multicolumn{15}{c}{\rule{0pt}{2ex} \textit{Setting 3: Using INF-X-7B Retriever} \rule{0pt}{2ex}} \\
        \midrule
        Retriever & INF-X (7B) & 63.4 & 79.8 & 70.9 & 69.9 & 73.3 & 57.7 & 64.3 & 61.9 & 56.1 & \textbf{54.5} & \textbf{51.9} & 53.1 & 67.9 \\
        Pointwise & ERank (32B) & 63.9 & 80.3 & \textbf{71.5} & 69.4 & 73.4 & 59.0 & 64.4 & 61.7 & \textbf{57.6} & 53.3 & 51.1 & 53.9 & 70.6 \\
        Listwise & ReasonRank (32B) & 45.3 & 60.2 & 54.2 & 48.5 & 60.8 & 44.2 & 39.5 & 47.0 & 33.4 & 37.0 & 20.3 & 45.5 & 53.0 \\
        Groupwise & GroupRank (32B) & \textbf{64.9} & \textbf{82.2} & 71.1 & \textbf{72.7} & \textbf{76.2} & \textbf{60.6} & \textbf{66.2} & \textbf{64.7} & 55.9 & 52.9 & 50.5 & \textbf{55.0} & \textbf{71.2} \\
        \bottomrule
    \end{tabular}}
\end{table*}

\subsection{Theoretical Efficiency} 
\label{sec:theoretical_efficiency}
\cref{tab:theoretical_efficiency} summarizes the theoretical efficiency of GroupRank against existing reranking paradigms based on three dimensions: generation requirements, parallel inference support, and computational complexity. As demonstrated, GroupRank achieves an optimal balance between low complexity and high practical throughput.While sorting-based methods like Setwise or Pairwise-heapsort offer competitive complexity, they are inherently limited by sequential dependencies that prevent effective parallel batching. In contrast, GroupRank supports full parallel inference, which is vital for minimizing end-to-end latency in real-world applications. Compared to pointwise methods that also allow batch processing, GroupRank reduces the number of required model forward passes from $\mathcal{O}(N)$ to $\mathcal{O}(\frac{N}{c})$ by processing multiple candidates simultaneously within a single prompt. Furthermore, our approach avoids the quadratic computational overhead associated with all-pair comparisons. By combining a reduced number of forward passes with the advantages of batch execution, GroupRank provides a high-throughput solution that remains scalable for large-scale applications.

\subsection{Robustness to Different First-stage Retrievers}
\cref{tab:retriever_comparison_v2} confirms GroupRank's robustness, as it \textbf{achieves the highest average performance across all three settings}. A critical observation is the shifting effectiveness of different paradigms as retrievers improve. While listwise methods perform well with BM25, they struggle with stronger baselines. In Setting 3, the listwise ReasonRank (45.3) falls drastically behind the INF-X retriever (63.4), whereas GroupRank (64.9) continues to provide positive gains. \textbf{This discrepancy stems from the output format}: listwise models generate discrete rankings that are difficult to integrate with first-stage scores through weighted fusion. Conversely, GroupRank's score-based outputs allow for seamless integration with retrieval priors. This flexibility ensures that GroupRank effectively leverages high-quality retrieval signals while maintaining the accuracy benefits of joint document scoring.

\section{Case Study}
To better understand the reasoning process of our GroupRank, we present the reasoning chains of our GroupRank(32B) on several BRIGHT datasets, including Biology (\cref{tab:case_biology}), Sustainable living (\cref{tab:case_sustainable_living}), Psycology (\cref{tab:case_psychology}), Robotics (\cref{tab:case_robotics}).

\onecolumn

\begin{promptbox}[prompt:pointwise_labeling]{Prompt of Pointwise Labeling using Qwen3-235B-instruct}
Your task is to rate how relevant and useful the document is for the query.
A document is **relevant and useful** if its content directly helps answer or address the query. A document is **not relevant or useful** if it does not provide content that helps answer the query, even if it mentions similar topics.
The answer should be 'Relevance score: X.' where X is a number from 0-10. 0 means completely irrelevant, and 10 means highly relevant and provides a complete, useful answer.

Here is the query:
{query}

Here is the document:
{passages}

Note that your answer must ONLY be in the format 'Relevance score: X', where X is a number from 0-10. Don't output anything else. 
\end{promptbox}

\begin{promptbox}[prompt:listwise_labeling]{Prompt of Listwise Labeling using Gemini2.5-Pro}
You are an expert passage reranker. Your task is to rank the provided passages based on how well they address the user's query, considering both **relevance and usefulness**.
Follow these steps:
1. **Understand the Query:** Identify the core question or intent behind the user's query.
2. **Evaluate Passages:** Think step-by-step to assess each passage. A passage is **valuable** if it directly and effectively helps answer the query. It is **not valuable** if it merely discusses similar topics without providing a direct answer.
3. **Rank & Output:**
- First, briefly explain your reasoning process for the ranking.
- Then, output a single JSON array containing the integer IDs of **all** provided passages. The array must be sorted from the most valuable passage to the least valuable.

The final output should look like this:
<Your reasoning here>
```json 
[1,2,...]
```

The user's query is:
{query}

Here are the passages to evaluate:
{passages_list} 
\end{promptbox}

\newpage
\begin{promptbox}[prompt:groupwise]{Prompt of Groupwise Reranking}
Your task is to evaluate and rank documents based on how well they help answer the given query. Follow this evaluation priority:
1. PRIMARY: Usefulness & Helpfulness - Does the document provide actionable information, solutions, or direct answers that help address the user's needs?
2. SECONDARY: Relevance - Does the document contain information related to the query topic?

Evaluation Process:
1. First, identify the user's core intent and what kind of help they need from the query
2. For each document, assess:
   - How directly it addresses the user's intent
   - What actionable information or answers it provides
   - How much it helps solve the user's problem or need
3. Compare documents against each other to ensure proper ranking
4. Assign scores that reflect the relative usefulness ranking

Scoring Scale (0-10):
- 9-10: Extremely helpful, directly answers the query with actionable information
- 7-8: Very helpful, provides substantial useful information for the query
- 5-6: Moderately helpful, contains some useful information but incomplete
- 3-4: Minimally helpful, limited useful information despite topic relevance
- 1-2: Barely helpful, mentions related topics but provides little useful information
- 0: Not helpful at all, cannot assist with answering the query

I will provide you {TOPK} documents, each indicated by a numerical identifier []. Score these documents based on their Usefulness and Relevance to the query.
Query: {QUERY}
Documents: {PASSAGES}

## Final Output Format
You must structure your response in exactly two parts: provide your brief reasoning process first, then output final scores in JSON format like below, with document IDs as string keys and integer scores as values for all {TOPK} documents. 
The reasoning process and answer are enclosed within <reason> </reason> and <answer> </answer> tags, respectively. Do NOT output anything outside the specified tags. Follow this exact format:
<reason> 
Analyze each document's usefulness and relevance to the query, explaining your scoring rationale
</reason> 
<answer> 
```json
{"[1]": 5, "[2]": 3, "[3]": 8, ...}
``` 
</answer>  
\end{promptbox}

\newpage
\begin{CaseStudyTable}{Case on Biology.}{tab:case_biology}
\textbf{Query} \\
\midrule
Claim in article about why insects are attracted to light. In this article they are addressing the reason insects are attracted to light when they say
Heat radiation as an attractive component is refuted by the effect of LED lighting, which supplies negligible infrared radiation yet still entraps vast numbers of insects.
I don't see why attraction to LEDs shows they're not seeking heat. Could they for example be evolutionarily programmed to associate light with heat? So that even though they don't encounter heat near/on the LEDs they still "expect" to?
\\
\midrule

\textbf{GroupRank Output}\\
\midrule
\fboxtag{<reason>}
The user is questioning an article's claim that insects are not attracted to light for its heat component, citing LED lighting as evidence. The user proposes an alternative hypothesis: insects might be evolutionarily programmed to associate light with heat, even if LEDs don't emit significant heat. Therefore, the most helpful documents will either directly address insect attraction to light, discuss evolutionary programming or learning related to light, or provide strong counter-evidence to the article's claim.\\ 
Here's the evaluation of each document:\\ 

[1]: Discusses phototropism in plants. While it talks about light attraction, it's not about insects, making it minimally helpful for the specific query.[2] Describes compound eyes in nocturnal insects and their ability to detect light. This is relevant to how insects perceive light but doesn't explain why they are attracted to it or address the heat aspect.[3] Contains only citation information, providing no content relevant to the query.[4] Discusses human color perception and metamerism. This is irrelevant to insect attraction to light or heat.\\ 

[5]Explains the molecular mechanism of light activation in cells (TRP channels). While it's about light perception, it's too general and doesn't connect to insect attraction or the heat hypothesis.[6] Focuses on circadian rhythms and light perception in Drosophila (fruit flies). This is highly relevant as it discusses how insects perceive and respond to light, which is foundational to understanding their attraction.[7] Discusses circadian rhythmicity and photoperiodism in animals, including their role in seasonal timing. This is relevant to how light influences animal behavior, which could include insects, but it's broader than specific attraction mechanisms.[8]: Describes bioluminescence in various insects and its use for attracting mates or luring prey. This directly illustrates how insects use light for attraction, supporting the idea of light being a significant stimulus, even if not for heat.[9]: Continues the discussion on circadian rhythms and photoperiodism, emphasizing the predictive nature of day length for physiology and behavior. This reinforces the importance of light cues in animal biology, which is relevant to insect attraction.\\  

[10]: Defines key mechanisms (photokinesis, phototaxis) of how organisms move towards light, making it central to the query.
[11]: Focuses on human circadian rhythms (melatonin, zeitgeber). Irrelevant as it does not address insects or the heat hypothesis.
[12]: Describes the physical structure of insect eyes (detection). Tangential as it doesn't explain the motivation for attraction.
[13]: Discusses the genetic basis of light perception in Drosophila (clock genes), providing a foundational biological explanation for attraction. Highly relevant.
[14]: Explains the physical link between heat and light color (Wien's law), directly supporting the user's core hypothesis. Extremely helpful.
[15]: Another document on the mechanics of insect vision (detection), not the underlying reason for attraction.
[16]: Discusses associative learning in plants, not insects. While conceptually related, the subject matter is incorrect, making it minimally relevant.\\ 

[17]: Citation only, with no content relevant to the query.
[18]: Same as [14], explaining Wien’s displacement law; highly relevant for the same reasons.
[19]: Describes diurnality/nocturnality and the role of light in animal activity and vision, giving broad evolutionary context for light’s importance, including for insects.
[20]: Covers circadian rhythms in animals (e.g., butterflies, moths) and sun-compass navigation; relevant to insect light perception but only indirectly to attraction to artificial light.\\

\textbf{Ranking Rationale}:\\
Documents [14] and [18] are the most helpful because they provide the scientific basis for the user's proposed hypothesis (association of light with heat via Wien's displacement law). Documents [6], [8], [10], and [13] are very helpful as they explain the mechanisms and specific examples of insect attraction to light, directly addressing the core phenomenon. Documents [7], [9], and [19] provide excellent contextual background on the importance of light and circadian rhythms in animal/insect biology. The remaining documents are less directly relevant, focusing on detection, human perception, or plant behavior, or are citations. 
\fboxtag{</reason>} \\

\fboxtag{<answer>}  \\ [2pt] 
{"[1]": 3, "[2]": 4, "[3]": 0, "[4]": 1, "[5]": 3, "[6]": 8, "[7]": 7, "[8]": 8, "[9]": 7, "[10]": 8, "[11]": 4, "[12]": 4, "[13]": 8, "[14]": 9, "[15]": 4, "[16]": 3, "[17]": 0, "[18]": 9, "[19]": 7, "[20]": 6}\\
\fboxtag{</answer>} \\
\end{CaseStudyTable}

\begin{CaseStudyTable}{Case on Sustainable Living.}{tab:case_sustainable_living}
\textbf{Query} \\
\midrule
How to recognize products with neonicotinoid pesticides in them? Recently, the EU has temporarily banned neonicotinoid pesticides because there are strong indications that these pesticides are responsible for the decline in bee populations. I've heard that this pesticide is not only used in bug sprays, but also in seemingly innocent products like flower bulbs, plant plugs and certain types of compost. Is this true? If so, how do I make sure I don’t already have this stuff at home? How can I recognize products or brands that contain/use these pesticides?
\\
\midrule

\textbf{GroupRank Output}\\
\midrule
\fboxtag{<reason>}
The user is asking how to recognize products containing neonicotinoid pesticides, specifically in items like flower bulbs, plant plugs, and compost. They also want to know if this is true and how to avoid them at home. I will prioritize documents that confirm the presence of neonics in these products, explain their persistence, and offer actionable advice on how to identify or avoid them. \\
 \\
 
[1]: This document discusses PFAS in compostable food packaging and biosolids. It is completely irrelevant to neonicotinoid pesticides. [2]: This is a list of product categories, including "Bulbs" and "Soil and Fertilizers." While it lists relevant product types, it provides no information about neonicotinoids or how to identify them. It's barely helpful as a list of product types. [3]: This document discusses composting dairy products. It is irrelevant to neonicotinoids. [4]: This document mentions concerns about pesticides in straw and the desire to find organic sources. While it doesn't name neonics, it highlights the general problem of pesticides in garden products and the need for organic alternatives, which is a relevant strategy for avoiding neonics. [5]: This document discusses horses in cities and electric carts. It is completely irrelevant. \\

[6]: This document is highly relevant. It explains that neonics remain active in soil for years and contaminate water, directly addressing the user's concern about their persistence and environmental impact. This helps explain *why* one needs to be careful with products that might contain them. [7]: This document lists various carbamate pesticides. It is irrelevant to neonicotinoids. [8]: This document discusses moisture content in compost piles. It is irrelevant to neonicotinoids. [9]: This document discusses composting in place and mentions avoiding cooked food, meat, grains, or dairy. It is irrelevant to neonicotinoids. [10]: This document discusses pottery glaze chemistry. It is completely irrelevant. \\

[11]: This document discusses the widespread exposure to neonicotinoids and links them to various health effects. While it doesn't directly help identify products, it reinforces the user's concern about avoiding them, making it moderately helpful for context. [12]: This document discusses synthetic pyrethroid barrier treatments for mosquitoes. It is irrelevant to neonicotinoids. [13]: This document lists various certifications like FSC, GREENGUARD, and Oeko-Tex. While these are general certifications for products, they do not specifically address neonicotinoids or pesticides in garden products. It's minimally helpful as a general concept of certification, but not specific enough. [14]: This document discusses plant-derived pyrethrins for mosquito control. It is irrelevant to neonicotinoids. [15]: This document is extremely helpful. It provides actionable information by directing the user to the EPA for pesticide registration and the National Pesticide Information Center (NPIC) for information on human health and insecticides. This is a direct way to find out what products contain neonics. \\

[16]: This document is very helpful. It confirms that neonicotinoids are used as seed treatments, which directly answers the user's question about their presence in "seemingly innocent products" like seeds (and by extension, plants grown from them). This is a key piece of information for recognition. [17]: This document is a personal anecdote with no relevance to the query. [18]: This document discusses global warming. It is completely irrelevant. [19]: This document discusses the properties of pure compost. It is irrelevant to neonicotinoids. [20]: This document is a table of contents for a Wikipedia page on neonicotinoids. It lists sections like "Agricultural usage," "Regulation," "Environmental and species impact," and "Bees." This confirms the user's premise about their use and impact, and suggests where to find more detailed information, making it very helpful for context and further research. 
\fboxtag{</reason>} \\

\fboxtag{<answer>}  \\ [2pt] 
\{"[1]": 0, "[2]": 1, "[3]": 0, "[4]": 4, "[5]": 0, "[6]": 8, "[7]": 0, "[8]": 0, "[9]": 0, "[10]": 0, "[11]": 5, "[12]": 0, "[13]": 2, "[14]": 0, "[15]": 9, "[16]": 8, "[17]": 0, "[18]": 0, "[19]": 0, "[20]": 7\} \\
\fboxtag{</answer>} \\
\end{CaseStudyTable}

\begin{CaseStudyTable}{Case on Pyschology.}{tab:case_psychology}
\textbf{Query} \\
\midrule
Can our beliefs change without reassessment or new evidence?
I think the strength of our beliefs can change, say from 90-95\% as the day wore on. But do people, people with or without psychosis, start and stop believing things without reasons?
Suppose when I stop and think I always know something is impossible. Does that mean, even in the throes of some irrational behaviour (I sleep with my cheating wife again), I know it is, at least without encountering new evidence to the contrary?
\\
\midrule

\textbf{GroupRank Output}\\
\midrule
\fboxtag{<reason>}
The user's query asks whether beliefs can change without reassessment or new evidence, specifically questioning if people can start and stop believing things without reasons, and if irrational behavior implies a lack of knowledge of impossibility. The core intent is to understand the mechanisms of belief change, particularly those that seem spontaneous or subconscious.\\ 
Here's a breakdown of how each document addresses the query: [1]: This document discusses the "mere-exposure effect" and how affective responses can precede cognitive ones, suggesting that preferences (and by extension, beliefs) can form or change without explicit conscious inference. This directly speaks to the idea of belief change without reassessment or new evidence, making it highly relevant.   [2]: This document describes hindsight bias and how people reconstruct memory to make past events seem inevitable, often by selectively retrieving information that supports the outcome. This is a clear example of belief change (about past events) occurring without new evidence, driven by cognitive distortions and the need for coherence.   [3]: This document defines "belief" as a non-sensory cognitive propositional attitude that represents the world and aims at truth. It explains that beliefs are formed or changed by perceptions and thoughts. While foundational, it doesn't directly address unconscious or reasonless belief change, but rather the general process.   [4]: This document discusses the emotional impact of realizing a belief is wrong due to new information. It focuses on the consequences of belief change, not the mechanisms of change itself, especially without evidence. \\

[5]: This document distinguishes factive and non-factive attitudes, defining belief as an attitude of assent towards a proposition, whether true or false. It clarifies that belief doesn't entail truth, which is relevant to the user's question about knowing something is impossible while still acting on it. It provides conceptual clarity but not direct mechanisms of change.   [6]: This is a reference to a paper title, not content, so it cannot be evaluated for usefulness. [7]: This document explains how schemas can alter memory and interpretation, leading to biased recollections. This is another mechanism where beliefs about past events or information can change without new evidence, driven by existing cognitive frameworks.   [8]: This document uses the "stop sign" example to illustrate that different emotional and behavioral reactions to the same event stem from different underlying beliefs. It highlights how beliefs shape perception and action, but doesn't explicitly discuss spontaneous belief change without reassessment.   \\

[9]: This document discusses motivated forgetting and repression, where unwanted memories or feelings are suppressed or destroyed. This is a mechanism by which beliefs or aspects of them can be altered or lost without conscious reassessment or new evidence, driven by psychological defense mechanisms.   [10]: This is a reference to a paper title, not content, so it cannot be evaluated for usefulness.   [11]: This document delves into social neuroscience, discussing how implicit attitudes and stereotypes can be activated and how conscious processes (like reappraisal) can modulate them. This is highly relevant as it describes how subconscious beliefs can influence perceptions and actions, and how they can be consciously managed, touching on both unconscious influence and conscious control.[12]: This document explains the "illusion of control," where people feel they control outcomes even when they don't, due to "skill cues" or a need to reassert control. This is a prime example of a belief (about control) forming or persisting without rational evidence, driven by psychological needs or heuristics.   [13]: This is a reference to a paper title, not content, so it cannot be evaluated for usefulness.   [14]: This document lists references for hindsight bias, indicating its existence and importance but providing no content itself.   [15]: This document elaborates on hindsight bias, comparing it to other memory distortions like the misinformation effect and false autobiographical memory. It details how these distortions can lead to changes in memory and belief without new evidence, driven by post-event information or internal processes.\\

[16]: States the core principle that memory is reconstructive and unreliable, providing foundational support for belief change without evidence.
[17]: Explains the "habit loop," a direct mechanism for how belief-like patterns are formed and reinforced automatically, without conscious reassessment.
[18]: Describes "emotional reasoning" as a mechanism where emotions override logic, causing belief changes that bypass rational thought.
\\

[19]: Lists several cognitive biases and effects (e.g., mere exposure, placebo) but only briefly; offers examples of belief-related phenomena without explaining mechanisms.
[20]: Extends [11] using social neuroscience and the distinction between declarative and nondeclarative memory, showing how different brain systems support explicit and implicit evaluations and subconscious belief processes.\\ 

Ranking Logic:
Documents that directly describe mechanisms of belief change without new evidence or reassessment, or that provide strong examples of such changes, are scored highest. Documents offering foundational definitions or general principles are moderately helpful. Documents that are references or too tangential are scored lowest.
\fboxtag{</reason>} \\

\fboxtag{<answer>}  \\ [2pt] 
{"[1]": 8, "[2]": 7, "[3]": 5, "[4]": 3, "[5]": 6, "[6]": 0, "[7]": 7, "[8]": 5, "[9]": 7, "[10]": 0, "[11]": 8, "[12]": 8, "[13]": 0, "[14]": 0, "[15]": 8, "[16]": 7, "[17]": 8, "[18]": 8, "[19]": 5, "[20]": 7} \\
\fboxtag{</answer>} \\
\end{CaseStudyTable}

\begin{CaseStudyTable}{Case on Robotics.}{tab:case_robotics}
\textbf{Query} \\
\midrule
behaviortree-cpp not installing using apt

I tried to install behaviortree using sudo apt install "ros-noetic-behaviortree-cpp" and got the following error while the system installing it:

Preparing to unpack \\
.../ros-noetic-behaviortree-cpp-v3\_3.8.5-1focal.20230814.151
459\_amd64.deb ... \\
Unpacking ros-noetic-behaviortree-cpp-v3\_3.8.5-1focal.20230814.151459 ...   \\
dpkg: error processing archive /var/cache/apt/archives/ros-noetic-behaviortree-c \\
pp-v3\_3.8.5-1focal.20230814.151459\_amd64.deb (--unpack): \\

trying to overwrite "/opt/ros/noetic/bin/bt3\_log\_cat", which is also in package "ros-noetic-behaviortree-cpp" \\ 

4.4.2-1focal.20231128.201059 dpkg-deb: error: paste subprocess was killed by signal (Broken pipe) 
\\
Errors were encountered while processing:

/var/cache/apt/archives/ros-noetic-behaviortree-cpp-v3\_3.8.5-1focal.20230814.15 \\

1459\_amd64.deb \\

E: Sub-process /usr/bin/dpkg returned an error code (1)
I tried multiple ways to install it but was unsuccessful.
\\
\midrule
\textbf{GroupRank Output} \\
\midrule
\fboxtag{<reason>}
The user is encountering an error during the installation of "ros-noetic-behaviortree-cpp-v3" via "apt". The core issue is a conflict between "ros-noetic-behaviortree-cpp-v3" and "ros-noetic-behaviortree-cpp", specifically regarding the file 
"/opt/ros/noetic/bin/bt3\_log\_cat". The user needs help resolving this conflict or finding an alternative installation method.
 \\
Here's the evaluation of each document: \\

\textbf{[1], [3], [4], [6], [9], [11], [14], [15], [17], [18]}: These documents list releases of "behaviortree\_cpp\_v3" for various ROS distributions (dashing, eloquent, melodic, kinetic, lunar, noetic). While they confirm the existence of the package for Noetic, they do not address the specific installation conflict or provide solutions for it. They are minimally helpful for context but don't solve the problem. \\
\textbf{[2], [5], [7], [19]}: These documents provide version numbers for "behaviortree\_cpp\_v3" in Noetic or general release repositories. Similar to the previous group, they offer versioning details but no insight into resolving the installation conflict. \\
\textbf{[8], [10], [12], [20]}: These documents mention "ros-noetic-behaviortree-cpp-v3\_3.8.6-1\_focal" and "debian/noetic/focal/behaviortree\_cpp\_v3". This indicates that a newer version (3.8.6) exists for Noetic, which might resolve the conflict if the user's current installation is outdated or if the conflict is due to a known bug in an older version. It suggests checking for updates or looking for a different package name/version, which could be a step towards resolution. \\
\textbf{[13]}: This document is highly relevant because it explains "dpkg" options like "--force-overwrite" and "--force-conflicts". The user's error message explicitly states trying to overwrite "/opt/ros/noetic/bin/bt3\_log\_cat", which is also in package "ros-noetic-behaviortree-cpp". This document provides actionable information on how to force dpkg to handle such conflicts, offering a potential workaround or solution to the immediate installation problem. \\
\textbf{[16]}: This document confirms that "behaviortree\_cpp\_v3" was released into the "noetic" distro. This is relevant for confirming the availability of the package for the user's ROS distribution, but it doesn't directly address the installation conflict. \\

Based on this, Document [13] is the most helpful as it offers a direct solution to the "dpkg" conflict. Documents [8], [10], [12], [20] are moderately helpful by suggesting newer versions or paths that might lead to a resolution. The rest are either too generic or only provide background information. 
\fboxtag{</reason>} \\

\fboxtag{<answer>}  \\ [2pt] 
\{"[1]": 3, "[2]": 3, "[3]": 3, "[4]": 3, "[5]": 3, "[6]": 3, "[7]": 3, "[8]": 5, "[9]": 3, "[10]": 5, "[11]": 3, "[12]": 5, "[13]": 9, "[14]": 3, "[15]": 3, "[16]": 4, "[17]": 3, "[18]": 3, "[19]": 3, "[20]": 5\} \\
\fboxtag{</answer>} \\
\end{CaseStudyTable}

\end{document}